\documentclass[superscriptaddress,twocolumn,showpacs,preprintnumbers,amsmath,amssymb,aps]{revtex4}

\usepackage{graphicx}  
\usepackage{dcolumn}   
\usepackage{bm}       
\usepackage{hyperref}
\usepackage{verbatim} 
\usepackage{natbib}
\usepackage{epsfig}
\usepackage{epstopdf}
\usepackage{amsmath}
\usepackage{booktabs} 
\usepackage{color}
\usepackage[toc,page]{appendix}
\usepackage{placeins}
\usepackage{float}
\usepackage{multirow}
\usepackage[normalem]{ulem}
\usepackage{amsmath}
\usepackage{tabu}
\usepackage[utf8]{inputenc}
\usepackage{rotating}
\usepackage{multirow}
\usepackage[export]{adjustbox}
\usepackage{csquotes}
\usepackage{tabularx,ragged2e,booktabs}

\let\oldhat\hat
\renewcommand{\vec}[1]{\mathbf{#1}}
\renewcommand{\hat}[1]{\oldhat{\mathbf{#1}}}

\begin{document}

\title{Wetting boundaries for ternary high density ratio lattice Boltzmann method}

\author{Neeru Bala}
\affiliation{Department of Mathematics, Physics and Electrical Engineering, 
Northumbria University, Newcastle upon Tyne NE1 8ST, UK}
\author{Marianna Pepona}
\affiliation{Department of Physics, Durham University, Durham, DH1 3LE, UK}
\author{Ilya Karlin}
\email{karlin@lav.mavt.ethz.ch}
\affiliation{Department of Mechanical and Process Engineering, ETH Zurich, CH-8092 Zurich, Switzerland}
\author{Halim Kusumaatmaja}
\email{halim.kusumaatmaja@durham.ac.uk}
\affiliation{Department of Physics, Durham University, Durham, DH1 3LE, UK}
\author{Ciro Semprebon}
\email{ciro.semprebon@northumbria.ac.uk}
\affiliation{Department of Mathematics, Physics and Electrical Engineering, 
Northumbria University, Newcastle upon Tyne NE1 8ST, UK}

	\date{\today}
	
\begin{abstract}
		
We extend a recently proposed ternary free energy lattice Boltzmann 
model with high density contrast \cite{Wohrwag2018}, by incorporating 
wetting boundaries at solid walls.
The approaches are based on forcing and geometric schemes,
with implementations optimised for ternary (and more generally
higher order multicomponent) models. 
Advantages and disadvantages of each method are addressed by performing
both static and dynamic tests, including the capillary filling dynamics 
of a liquid displacing the gas phase, and the self-propelled motion 
of a train of drops.
Furthermore, we measure dynamic angles and show that the slip length 
critically depends on the equilibrium value of the contact angles, 
and whether it belongs to liquid-liquid or liquid-gas interfaces. 
These results validate the model capabilities of simulating complex ternary fluid dynamic problems near solid boundaries, for example 
drop impact solid substrates covered by a lubricant layer.
\end{abstract}
	
\maketitle


\section{Introduction}

Understanding the flow properties of ternary fluid systems is 
key in many natural phenomena and technological applications. 
In microfluidics, combinations of immiscible liquids are
employed to produce multiple emulsions \cite{Utada2005}.
In food sciences and pharmacology, collisions between immiscible 
drops \cite{Planchette2011} and liquid streams \cite{Planchette2018} 
can be exploited to encapsulate liquids. 
Collisions are also particularly relevant in combustion
engines, where encapsulation of water drops by fuel can 
induce micro-explosions enhancing the burning rate \cite{Wang2004a}.
Furthermore, in advanced oil recovery, the Water-Alternate-Gas (WAG) 
techniques are frequently employed to enhance the recovery \cite{Changlin-lin2013}.
The oil-water interaction in dynamic conditions also poses 
environmental challenges. For example the spilling of an oil layer 
at the surface of sea water strongly affects the production of 
marine aerosol when rain drops impact on the oil layer \cite{Murphy2015}.
In contrast, placing a lubricant layer on a rough solid is the
key idea behind the recent development of Slippery Lubricant 
Impregnated Surfaces (SLIPS), allowing to virtually eliminate 
contact line pinning \cite{Smith2013,Keiser2017}, 
with applications in coatings and packaging.

Several numerical schemes have been proposed in the recent years 
to simulate ternary and higher order fluid systems,
including immersed boundary \cite{Hua2014a}, level set \cite{Merriman1994} 
and phase field methods \cite{Gao2011,Kim2012b,Zhang2016a,HaghaniHassanAbadi2018,Park2018}. 
In this work we employ the Lattice Boltzmann (LB) method \cite{Succi2001,Kruger2017}. 
Multiphase and multicomponent LB  models are characterised
by a diffuse interface, which has the advantage that the interface does not need  
to be tracked explicitly \cite{Kusumaatmaja2006,Kusumaatmaja2007}. This makes LB models particularly convenient
to study problems involving coalescence or break-up of liquids
\cite{Ledesma-Aguilar2011,Komrakova2014}, 
drop collisions \cite{Premnath2005,MazloomiMoqaddam2015,Moqaddam2016,Montessori2017},
drop impact on solid walls \cite{Mukherjee2007,Lee2010,Banari2014,AshokeRaman2016,Liu2015c,Andrew2017,AshokeRaman2016} 
and on topographic or chemically patterned surfaces \cite{Sbragaglia2013,MazloomiMoqaddam2017a}.

Several LB models have been proposed to study ternary fluid systems.
Travasso et al. proposed a free energy model to study phase separation of ternary
mixtures under shear \cite{Travasso2005}; Spencer et al. proposed a color model
to study $N>2$ component systems \cite{Spencer2010}; 
Ridl et al. proposed a model
combining N Van der Waals equation of states to study the stability of multicomponent
mixtures \cite{Ridl2018}; Semprebon et al. proposed a ternary free energy 
approach \cite{Semprebon2016} to model liquids with equal density,
showing that the method can simulate drop morphologies on SLIPS \cite{Semprebon2017}
and their dynamic properties \cite{Sadullah2018} for a wide range 
of surface tensions and contact angles.

To model effects of inertia, the large density ratio between liquids 
and the gas phase needs to be accounted \cite{MazloomiM2015,Liang2017b},
but only recently ternary morels for high density ration have been proposed.
Shi et al. \cite{Shi2016} extended the binary Cahn-Hilliard model for high 
density ratio proposed by Want et al. \cite{Wang2015c} to three components.
W\"ohrwag et al proposed a free energy functional combining multiphase and
multicomponent terms \cite{Wohrwag2018}, and employing the entropic 
collision approach \cite{Chikatamarla2006} could simulate density contrast up to $10^3$.
The model could capture the salient features of head-on collisions between 
immiscible drops, reproducing the bouncing, adhesion and encapsulation mechanisms
previously observed experimentally\cite{Planchette2011}. 

In this work we extend this high density ternary approach to model
wetting of solid boundaries. The paper is organised as follows: 
In section \ref{model} we summarise the ternary
model introduced in Ref. \cite{Wohrwag2018}.
In section \ref{tensions} we perform an extensive analysis of 
the interfacial properties as function of the free energy parameters.
In section \ref{boundaries} we describe our  
implementation of three methods for wetting of solid boundaries, 
namely \emph{force}, \emph{geometric extrapolation} and
\emph{geometric interpolation}, and benchmark the accuracy
of contact angles in mechanical equilibrium.
In section \ref{filling} we compare the accuracy of the \emph{force}
and \emph{geometric interpolation} methods in simulating the 
capillary filling of a 2D channel.
In section \ref{bislug} we perform a ternary-specific benchmark,
simulating a self-propelled bi-slug.
This will enable us to evaluate the slip properties
of the three fluid interfaces and assess the impact of
different slip mechanisms.
Finally, in section \ref{conclusions} we summarise and discuss 
our results.


\section{Lattice Boltzmann formulation}
\label{model}

In this section, we summarise the derivation of the multiphase-multicomponent 
lattice Boltzmann model proposed in Ref. \cite{Wohrwag2018}.

\subsection{Ternary multiphase-multicomponent free energy} 

The ternary free energy model is conveniently expressed in terms
of a combination of bulk $f_{\rm Bulk}$ and the interfacial terms 
$f_{\rm Inter}$ in the free energy functional:
\begin{equation}
\label{equ:TotalEnergy}
F=\int\left[f_{Bulk} + f_{Inter} \right] dV,
\end{equation} 	
where		
\begin{eqnarray}
\label{equ:BulkEnergy}
f_{Bulk} &=& \frac{\lambda_1}{2} (\Psi_{\rm eos}(\rho) - \Psi_0) +  \\
&& \frac{\lambda_2}{2} C_{2}^2(1-C_{2})^2 +\frac{\lambda_3}{2} C_{3}^2(1-C_{3})^2, 
\nonumber \\
\label{equ:InterfEnergy}
f_{Inter} &=& \frac{\kappa_1}{2} (\vec{\nabla} \rho)^2 
+ \frac{\kappa_2}{2}(\vec{\nabla} C_{2})^2 
+\frac{\kappa_3}{2} (\vec{\nabla} C_{3})^2.
\end{eqnarray}

The first term in Eqn. \eqref{equ:BulkEnergy}  tunes the coexistence of high 
density ($\rho_l$) liquid with a low density ($\rho_g$) gas.
The term $\Psi_{eos}(\rho)$ is derived by integrating any suitable non ideal 
equation of state, $p_{eos}=\rho(d \Psi_{eos} / d \rho)-\Psi_{eos}$.
In our previous work \cite{Wohrwag2018} we have shown that the model 
can host various equations of state \cite{Yuan2006}, 
including van der Waals, Peng-Robinson and Carnahan-Starling.
Here we employ the Carnahan-Starling equation of state:
\begin{equation}
\label{equ:PsiEOS}
\Psi_{eos}=\rho\left(C-a\rho-\frac{8RT(-6+b\rho)}{(-4+b\rho)^2} +RT\log(\rho)\right),
\end{equation} 
where the constants $C$ and $\Psi_0$ enforce 
$\Psi_{eos}(\rho_g)= \Psi_{eos}(\rho_l)=\Psi_0$. This condition ensures 
that the common tangent construction is valid for all coexisting phases. 
Unless otherwise stated, we employ the following values 
$a=0.037$, $b=0.2$ and $R=1$, for which the critical 
temperature is $T_c=0.3373\frac{a}{bR}$.

We define the relative concentration of the gas phase as
\begin{equation}
\label{equ:definitionC1}
C_1= \frac{\rho-\rho_l}{\rho_g-\rho_l},
\end{equation} 
for which $C_1=0$ when $\rho = \rho_l$ and $C_1=1$ when $\rho = \rho_g$.

The second and third terms in Eqn. \eqref{equ:BulkEnergy} represent a double well potential, 
as function of the relative liquid concentrations: $C_{2}$ and $C_{3}$. 
Each concentration has two minima at $C_{2,3}=0$ and $C_{2,3}=1$ 
corresponding to the presence or absence of the liquid.
For convenience we introduce the phase field $\phi=\chi(C_2-C_3)$ which, together with the density
$\rho$, describes the system state. 
The parameter $\chi$ usually takes the value $\chi=5$ in our model,
and is employed to rescale the field $\phi$ such as the distance between
minima is similar in both the $\rho$ and $\phi$ fields.
The variable transformations
\begin{equation}
\label{equ:definitionC2}
C_{2} = \frac{1}{2} \left[1+\frac{\phi}{\chi} - \frac{\rho-\rho_l}{\rho_g-\rho_l}\right],
\end{equation} 
and
\begin{equation}
\label{equ:definitionC3}
C_{3} = \frac{1}{2} \left[1-\frac{\phi}{\chi}- \frac{\rho-\rho_l}{\rho_g-\rho_l}\right],
\end{equation} 
enforce the constraint 
\begin{equation}
\label{equ:definitionCsum}
C_{1} + C_{2} + C_{3} = 1
\end{equation} 
and allow us to map the density and phase field to the concentration fields. 

The bulk free energy density in Eqn. \eqref{equ:BulkEnergy} describes three distinct 
energy minima in the $(\rho ,\phi)$ space, corresponding to  $(\rho ,\phi) = (\rho_g,0)$ 
(gas phase), and $(\rho_l,+\chi)$, $(\rho_l,-\chi)$ (liquid phases). 
The set of lambdas ($\lambda_1$, $\lambda_2$ and $\lambda_3$) tunes the magnitude of the energy barriers between each pair of phases.

Eqn. \eqref{equ:InterfEnergy} contains gradient terms of the density field and
the concentration of the two liquid components, describing the energy penalty
in the formation of the interfaces, tuned by the set of kappas ($\kappa_1$, $\kappa_2$ and $\kappa_3$).
Summarising, this free energy model depends 
on six independent parameters, to fully determine the thermodynamic properties of the system. 
	
\subsection{Derivation of the pressure tensor} 	

The chemical potentials $\mu_\rho$ and $\mu_\phi$ are obtained 
directly from the free energy 
\begin{eqnarray}
\mu_{\rho}(\vec{r})= \frac{\delta F}{\delta \rho(\vec{r})} &=& \mu^{Bulk}_{\rho} + \mu^{Inter}_{\rho} \\
\mu_{\phi}(\vec{r})= \frac{\delta F}{\delta \phi(\vec{r})} &=&  \mu^{Bulk}_{\phi} + \mu^{Inter}_{\phi}.
\end{eqnarray}
For convenience we express the chemical potentials in terms of the relative 
concentrations and, to simplify the notation, we define the auxiliary function  
$ g(x) = x(x - 1/2)(x - 1)$:
\begin{eqnarray}
\mu^{Bulk}_{\rho} &=& \frac{\lambda_1}{2} \frac{d \Psi_{eos}}{d \rho}  
- \frac{\lambda_2 }{\Delta\rho} g(C_2)
+ \frac{\lambda_3 }{\Delta\rho} g(C_3), \label{equ:bulkchempotrho}  \\
\mu^{Bulk}_{\phi} &=& 
\frac{\lambda_2 }{\chi} g(C_2)
- \frac{\lambda_3 }{\chi} g(C_3), \label{equ:bulkchempotphi} \\
\mu^{Inter}_{\rho} &=& - \kappa_{\rho\rho} \nabla^2 \rho - \kappa_{\rho\phi} \nabla^2 \phi ,
\label{equ:gradchempotrho} \\
\mu^{Inter}_{\phi} &=& \kappa_{\rho\phi} \nabla^2 \rho - \kappa_{\phi\phi} \nabla^2 \phi.
\label{equ:gradchempotphi}
\end{eqnarray}

In Eqn. \eqref{equ:bulkchempotrho} $d\Psi_{eos}/d\rho$ is the first derivative by the density
of the non ideal equation of state, and $\Delta \rho=\rho_l -\rho_g$.
For the Carnahan-Starling EOS the first derivative of Eqn. \eqref{equ:PsiEOS} is
\begin{equation}
\frac{d\Psi_{eos}}{d\rho} = C - 2a\rho + RT(1+\log\rho) + \frac{16RT(b\rho-12)}{(-4+b\rho)^3}.
\end{equation}
Furthermore, in Eqns. \eqref{equ:gradchempotrho} and \eqref{equ:gradchempotphi} 
the mixing coefficients for the gradient terms are
\begin{eqnarray}
\kappa_{\rho\rho} &=& \left[ \kappa_1 + \frac{\kappa_2 + \kappa_3}{4(\rho_g-\rho_l)^2 }\right], \\
\kappa_{\phi\phi} &=& \frac{\kappa_2 + \kappa_3}{4\chi^2}, \\
\kappa_{\rho\phi} &=& -\frac{\kappa_3 - \kappa_2}{4\chi(\rho_g-\rho_l)}.
\end{eqnarray}
	
The pressure tensor can be inferred from the relation
$\nabla\cdot \vec{P}=\rho\nabla\mu_\rho + \phi\nabla\mu_\phi$
and takes the form
\begin{eqnarray}
\label{equ:pressuretensor}
P_{\alpha\beta} &=& p_0 \delta_{\alpha\beta} \\
&+& \kappa_{\rho\rho} \left[(\partial_\alpha \rho)(\partial_\beta \rho) 
- \left( \rho (\partial_{\gamma\gamma} \rho) 
+ \frac{1}{2}(\partial_\gamma \rho)^2\right) \delta_{\alpha\beta} \right] \nonumber \\
&+& \kappa_{\phi\phi} \left[(\partial_\alpha \phi)(\partial_\beta \phi) 
- \left( \phi (\partial_{\gamma\gamma} \phi) 
+ \frac{1}{2}(\partial_\gamma \phi)^2\right) \delta_{\alpha\beta} \right] \nonumber \\
&+& \kappa_{\rho\phi} \left[(\partial_\alpha \rho)(\partial_\beta \phi)
 + (\partial_\alpha \phi)(\partial_\beta \rho) \right. \nonumber \\
&& \left. - \left( \rho (\partial_{\gamma\gamma} \phi) 
+ \phi (\partial_{\gamma\gamma} \rho) 
+ (\partial_\gamma \rho)(\partial_\gamma \phi) \right) \delta_{\alpha\beta} \right], \nonumber
\end{eqnarray}
where $p_0$ is the pressure in the fluid bulk
\begin{equation}
\label{equ:bulkpressure}
p_0 = \rho \mu^{Bulk}_{\rho} + \phi \mu^{Bulk}_{\phi} - f_{Bulk}.
\end{equation}

\subsection{Entropic Lattice Boltzmann Implementation} 
	
The dynamic evolution of the isothermic ternary system follows the
continuity, Navier-Stokes, and Cahn-Hilliard equations:
\begin{equation}
\label{equ:Continuity}  
\partial_t \rho + \vec{\nabla} \cdot \left( \rho \vec{v} \right) = 0, 
\end{equation}
\begin{eqnarray}
\partial_t (\rho \vec{v}) + \vec{\nabla} \cdot \left( \rho \vec{v} 
\otimes \vec{v} \right) = \nonumber \\
- \vec{\nabla} \cdot \vec{P} + \vec{\nabla} 
\cdot [\eta ( \vec{\nabla v} +  \vec{\nabla v^T} )], 
\label{equ:NS}  
\end{eqnarray}
\begin{equation}
\partial_t  \phi + \vec{\nabla} \cdot (\phi \vec{v}) = M \nabla^2 \mu_\phi, 
\label{equ:Cahn-Hilliard} 
\end{equation}
where $\vec{v}$ is the fluid velocity, $\eta$ is the dynamic viscosity, and 
$M$ represents the mobility in the Cahn-Hilliard model for the order parameter $\phi$. 
	
To solve the equations of motion we introduce two sets of distribution functions, 
evolving the density $\rho$ and the order parameter $\phi$. 
For the density $\rho$, we employ the entropic lattice Boltzmann method (ELBM)
\cite{MazloomiM2015,M2015}, 
\begin{eqnarray}
\label{equ:ELBMequation}
 f_i(\vec{x}+\vec{c}_i \Delta t,t+\Delta t)  = \nonumber \\
 f_i(\vec{x},t) + \alpha\beta\left[ f_i^{eq}(\rho,\vec{u})-f_i(\vec{x},t)\right] + F_i.
\end{eqnarray}
We implement the exact form for the equilibrium distribution function 
$f_i^{eq}(\rho,\vec{u})$, which for a $D$-dimensional system (described
by the lattices D1Q3, D2Q9 or D3Q27) can be written 
in the product form \cite{Karlin1999,Ansumali2003},
\begin{equation}
\label{equ:eqrho}
f_i^{eq}=\rho w_i   \Pi_{\alpha=1}^{D}A(u_\alpha)[B(u_\alpha)]^{c_{i\alpha}}.  
\end{equation}
The $w_i$'s are the lattice weights and $c_{i\alpha}$ is the $\alpha$ component 
of the $\vec{c}_i$-th lattice vector. The functions $A(u)$ and $B(u)$ are given by 
\begin{equation}
\label{equ:A}
A(u)=2-\sqrt{1+3u^2},
\end{equation}
and
\begin{equation}
\label{equ:B}
B(u)=\frac{2u+\sqrt{1+3u^2}}{1-u}.
\end{equation}
	
The forcing term $F_i$ in Eqn. \eqref{equ:ELBMequation} \cite{MazloomiMoqaddam2015}	
is implemented via the Exact Differences scheme
\begin{equation}
\label{equ:ELBMForcingExactDiff}
F_i= \left[ f_i^{eq}(\rho,\vec{u}+\delta \vec{u})-f_i^{eq}(\rho,\vec{u}) \right],
\end{equation}
where $\rho \vec{u} = \sum_i f_i \vec{c}_i =  \sum_i f^{eq}_i \vec{c}_i$ 
is the bare fluid velocity, and $\delta\vec{u}=(\vec{F}/\rho)\Delta t$ is
the correction to the fluid velocity arising from the force
\begin{equation}
\label{equ:force}
\vec{F}=\vec{\nabla}\cdot(\rho c_s^2 \vec{I} -\vec{P} \bm).
\end{equation}
In Eqn. \eqref{equ:force} $c_s^2=1/3$ is the speed of sound in the lattice Boltzmann scheme.

In ELBM the parameter $\beta$ tunes the kinematic viscosity $\nu=\eta/\rho=(\beta^{-1}-1)/6$, 
and the parameter $\alpha$ is the non-trivial root of 
\begin{equation}
\label{equ:EntropyEquation}
H\left(f'+\alpha\left[f^{eq}(\rho,\vec{u}+\delta\vec{u})-f'\right]\right)=H\left(f'\right).
\end{equation}
In Eqn. \eqref{equ:EntropyEquation}  
\begin{equation}
\label{equ:mirror_state}
f'_i=f_i+[f_i^{eq}(\rho,\vec{u}+\delta\vec{u})-f_i^{eq}(\rho,\vec{u})]
\end{equation}
represents the mirror state, and 	
\begin{equation}
\label{equ:EntropyEstimate}
H(f) = \sum_i f_i \ln(f_i/w_i), 
\end{equation}
is the entropy.

To evolve the order parameter $\phi$, we employ a standard LBGK scheme 
\begin{equation}
\label{equ:CahnHilliardLBM}
g_i(\vec{x}+\vec{c}_i \Delta t,t+\Delta t) =  
g_i(\vec{x},t) + \frac{\left[g_i^{eq}(\phi,\vec{v})-g_i(\vec{x},t)\right]}{\tau}.
\end{equation}
The parameter $\tau$ is related to the mobility $M=\Gamma\left( \tau- 1/2 \right)$
in Eqn.\eqref{equ:Cahn-Hilliard}, where the constant $\Gamma$ tunes the diffusivity 
and is chosen to be $\Gamma=1$ unless otherwise stated. 
The equilibrium distribution function, $g_i^{eq}(\phi,\vec{v})$, can be written as
\begin{eqnarray}
g_i^{eq}(\phi,\vec{v}) &=& w_i \left( \frac{\Gamma \mu_\phi}{c_s^2} 
+ \frac{\phi v_\alpha c_{i\alpha}}{c_s^2} 
\right. \nonumber \\ && \left. 
+ \frac{\phi v_\alpha v_\beta 
\left(c_{i\alpha}c_{i\beta}-c_s^2\delta_{\alpha\beta}\right)}{2c_s^4}  \right), \\
g_0^{eq}(\phi,\vec{v}) &=& \phi - \sum_{i\neq 0} g_i^{eq}, 
\end{eqnarray}
where the actual fluid velocity, $\vec{v} = \vec{u} + \delta \vec{u}/2$,
is required.

\section{Surface tensions}
\label{tensions}

In free energy models based on double well potentials \cite{Briant2004a,Briant2004b}, 
the shape of the concentration profile against the spatial coordinates takes the form of a 
hyperbolic tangent. This feature is inherited in our ternary model, but only for 
the liquid-liquid interface between phases $C_2$ and $C_3$, which is characterised by 
the parameter 
\begin{equation}
\label{equ:liquidliquidalpha}
\alpha_{23}=\sqrt{\frac{\kappa_2+\kappa_3}{\lambda_2+\lambda_3}}.
\end{equation}
We can assume that the density does not vary at the interface between $C_2$ and $C_3$, 
and set $C_1=0$ along the interface. Following Ref. \cite{Semprebon2016}, 
if the coordinate $x$ measures the distance from the interface along its normal direction, 
the concentration profiles of the components 
$C_2$ and $C_3$ vary according to
\begin{equation}
\label{equ:profile}
C_{2,3}(x)=\frac{1 \pm \tanh{\frac{x}{2\alpha_{23}}}}{2}.
\end{equation}	
Integrating of the concentration profiles along $x$ we derive a simple expression
for the surface tension \cite{Semprebon2016}
\begin{equation}
\label{equ:tension23}
\gamma_{23}=\frac{\alpha_{23}}{6}(\lambda_2+\lambda_3).
\end{equation}	

For the liquid-gas interfaces, it is not possible to assume a priori that the
$\rho$ and $\phi$ fields vary in the same way.
Indeed, the minimization of the free energy seeks for a path which cannot
be described analytically.
To illustrate this aspect, we study in detail four cases, represented by the
parameter sets reported in table \ref{tab:tabletension}. 
For each set we independently compute for all interfaces the surface 
tension $\gamma=\Delta P R$ by measuring the pressure jump $\Delta P$ 
across the interface of 2D drop of radius $R$ (bubble test).

\begin{table}[tbh]
\begin{tabular}{|c||c|c|c|c| }
\hline
{\bf set  } & {\bf 1} & {\bf 2} &{\bf 3} &{\bf 4} \\
\hline
\hline
$\bm \lambda_1$ & $0.6$ & $0.6$ & $0.01$ & $0.1$ \\ 
$\bm \kappa_1$ & $0.01$ & $0.01$ & $0.01$ & $0.01$ \\
\hline
$\bm \lambda_2$ & $1.0$ & $1.1$ & $1.5$ & $1.0$  \\ 
$\bm \kappa_2$ & $1.0$ & $1.1$ & $1.5$ & $1.6$  \\ 
\hline
$\bm \lambda_3$ & $1.0$ & $0.5$ & $1.5$ & $0.2$  \\ 
$\bm \kappa_3$ & $1.0$ & $0.5$ & $1.5$ & $-0.4$  \\ 
\hline
$\bm \gamma_{12}$ & $0.414$ & $0.431$ & $0.333$ & $0.321$  \\ 
$\bm \gamma_{13}$ & $0.414$ & $0.334$ & $0.333$ & $0.120$  \\ 
$\bm \gamma_{23}$ & $0.323$ & $0.259$ & $0.485$ & $0.180$  \\ 
\hline
$\bm \theta_1$ & $134.1$ & $143.1$ & $86.3$ & $ - $  \\ 
$\bm \theta_2$ & $112.9$ & $129.2$ & $136.8$ & $ -$  \\ 
$\bm \theta_3$ & $112.9$ & $87.6$ & $136.8$ & $ - $  \\
\hline
&
\includegraphics[width=0.09\textwidth]{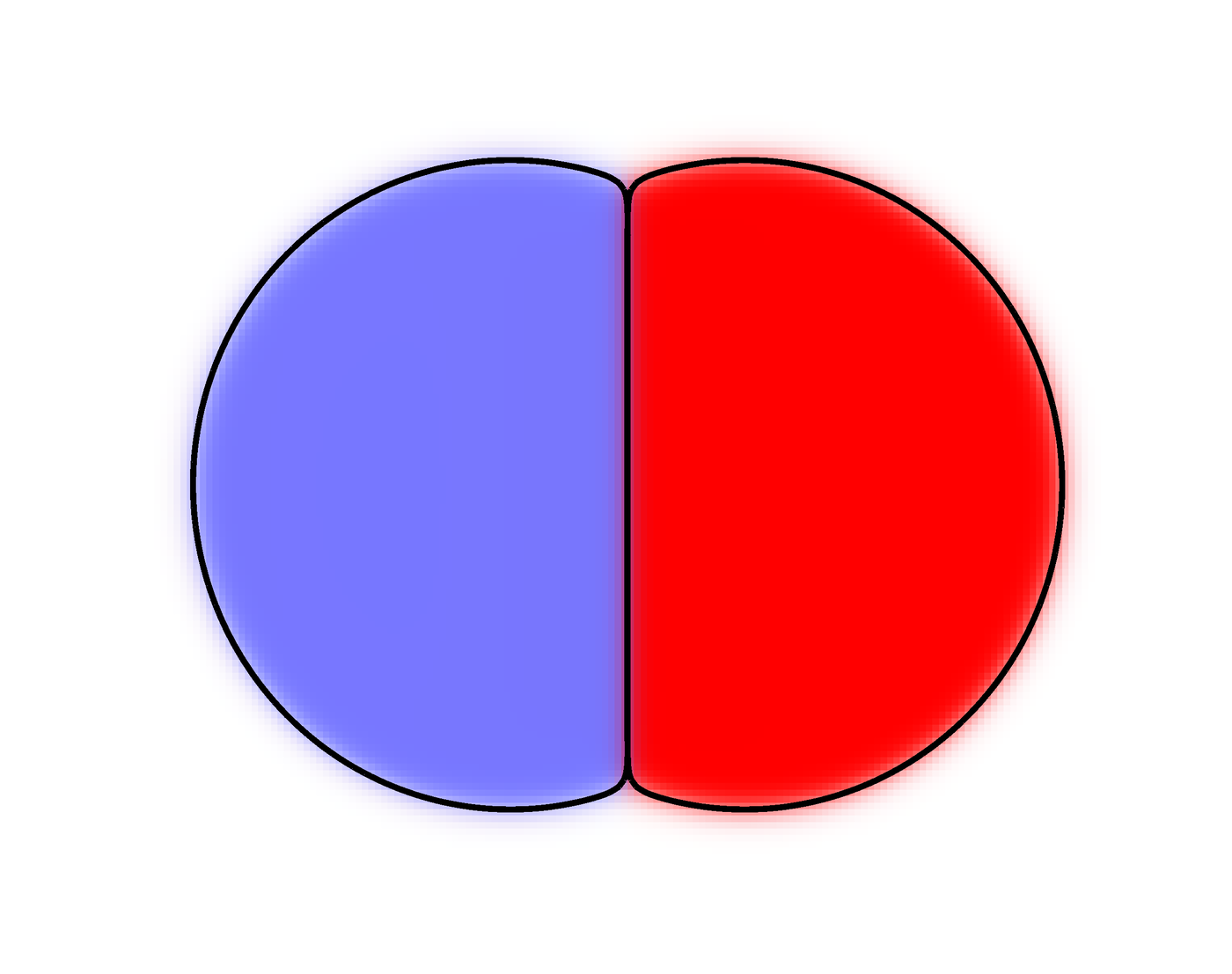}
&
\includegraphics[width=0.09\textwidth]{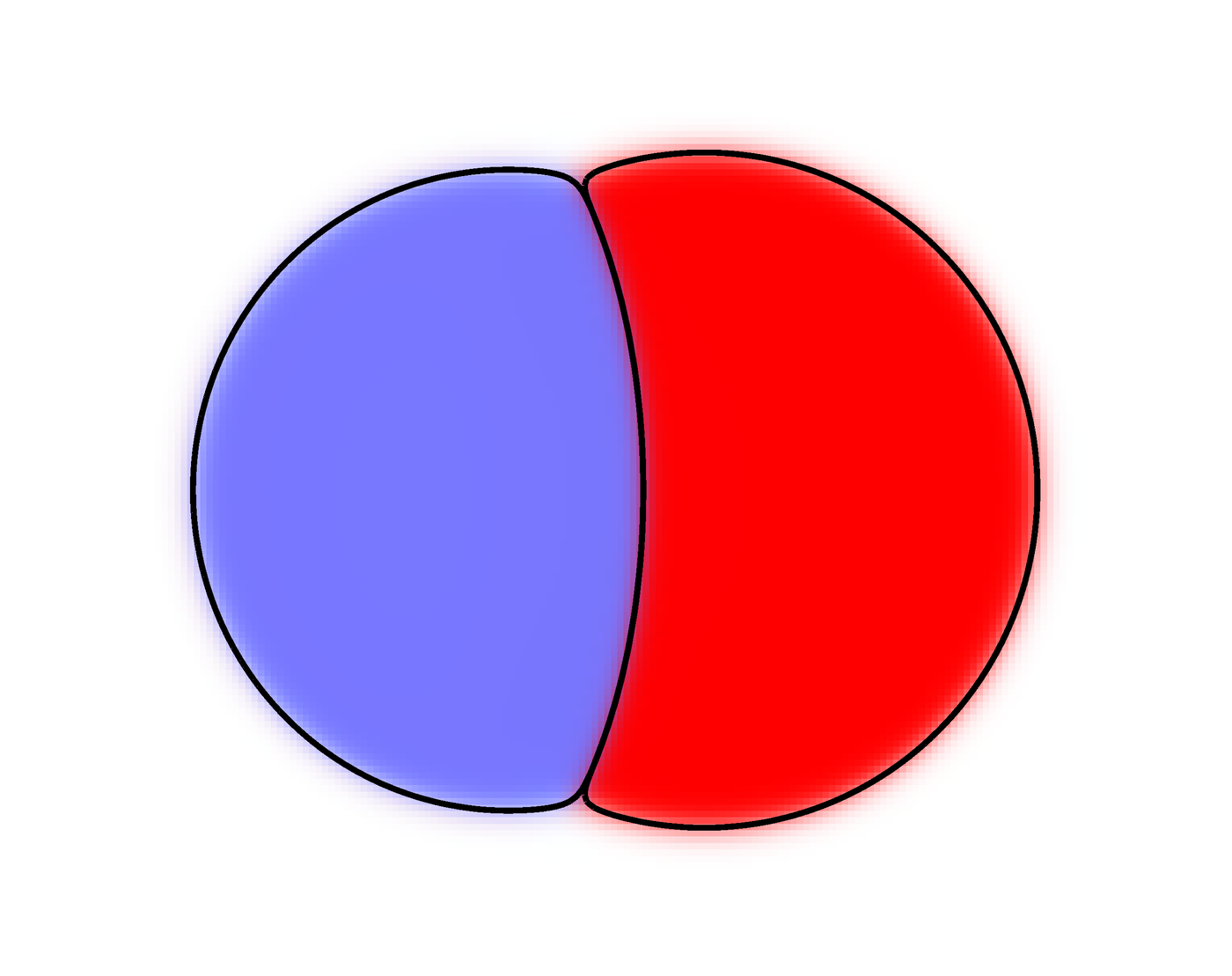}
&
\includegraphics[width=0.09\textwidth]{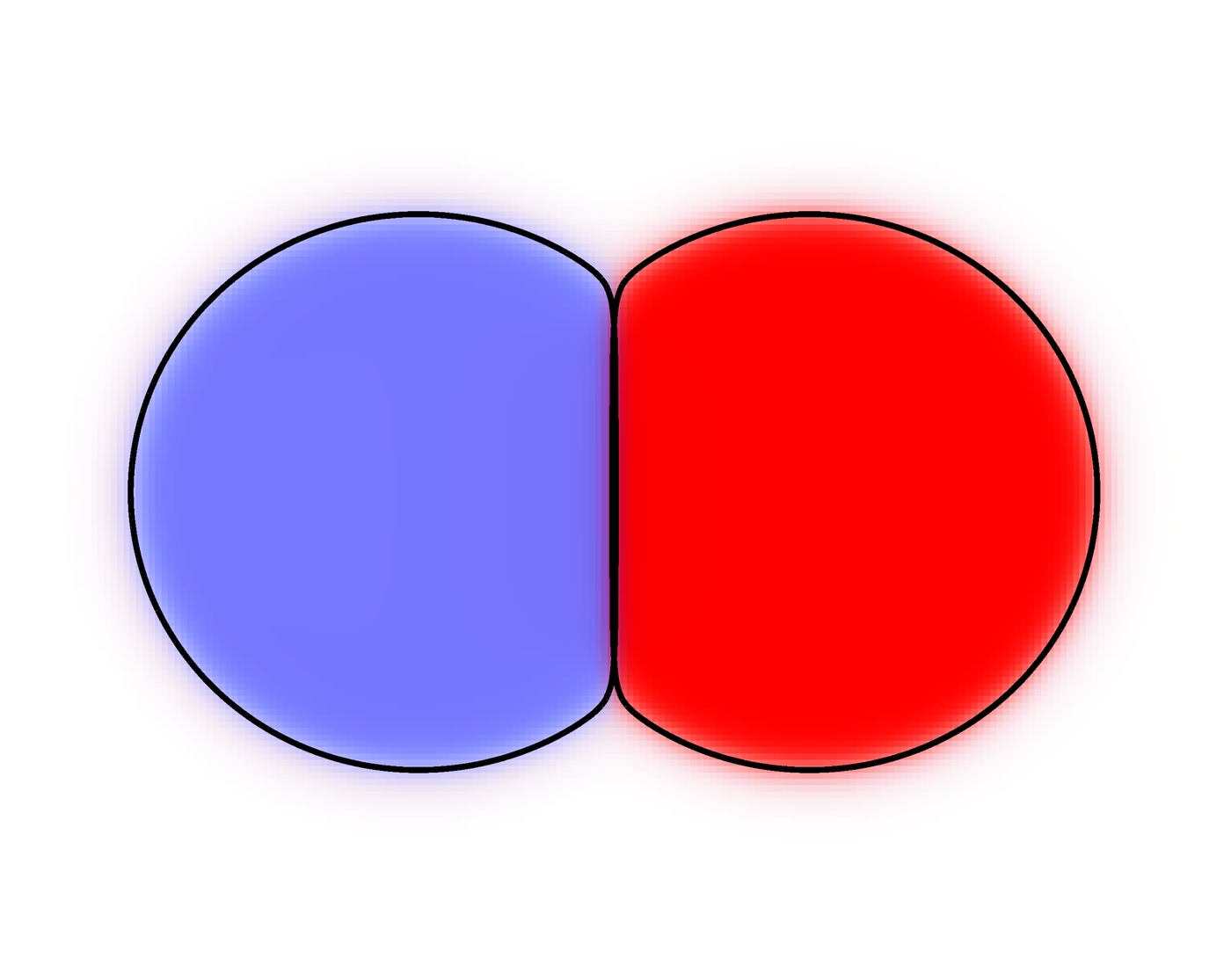}
&
\includegraphics[width=0.09\textwidth]{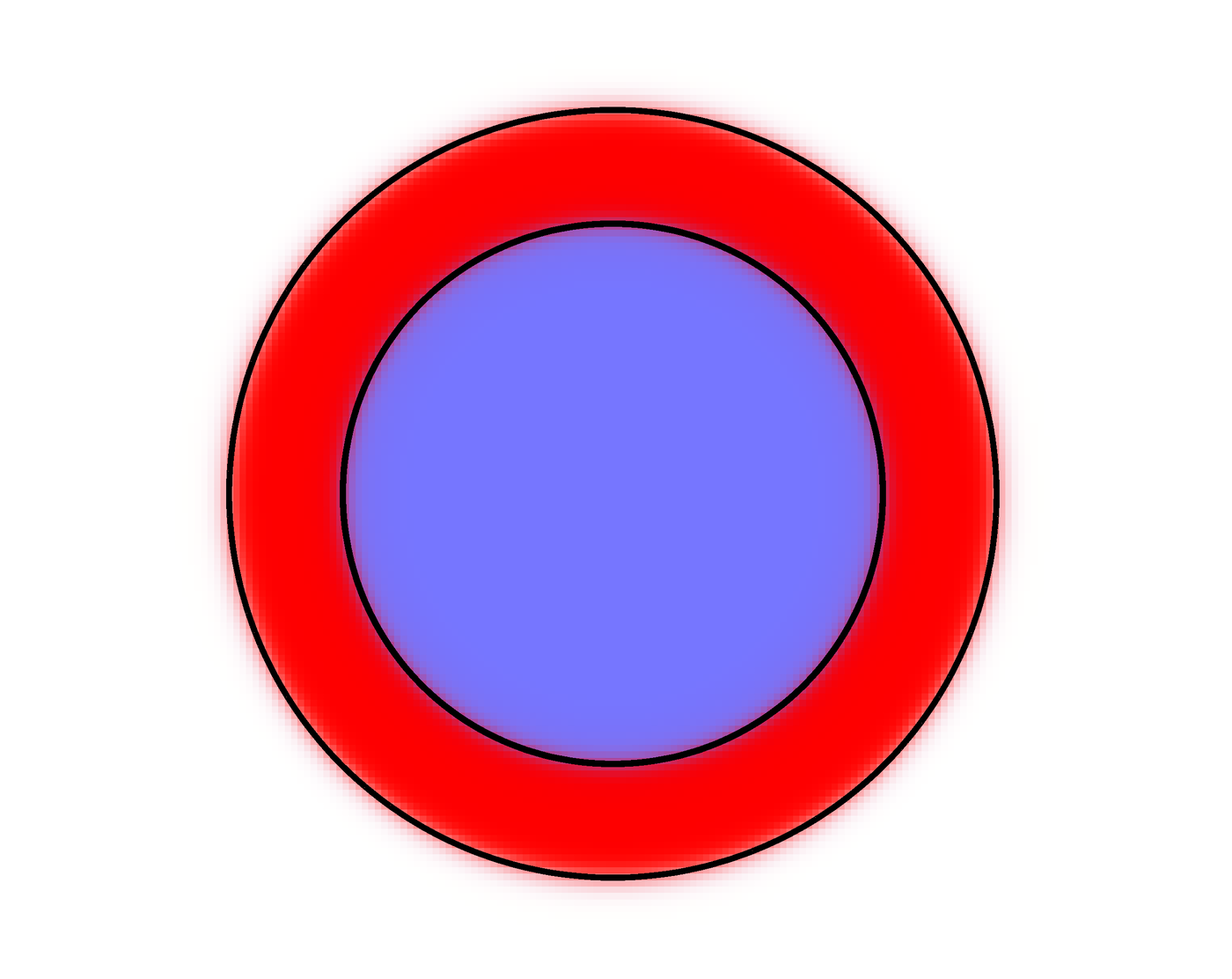}  \\
\hline
\end{tabular}
\caption{\label{tab:tabletension}
Parameters of four selected sets, and the relative surface tensions 
and Neumann angles. The last row reports the global energy minimum 
configuration of a double emulsion. The white region corresponds to 
the gas phase ($C_1$), while the blue and red regions correspond to 
the liquids $C_2$ and $C_3$.}
\end{table}

The 4 sets are listed in order of increasing mismatch between the interfacial profiles.
The first set represents two liquids with symmetric properties, where the
liquid-liquid surface tension is slightly lower than both the liquid-gas ones. 
The second set describes three fluids with different properties.
The third set also describes two equal liquids but $\lambda_1$ is much
smaller than in the first set, leading to a liquid-liquid 
surface tension significantly larger than the liquid-gas ones. 
The fourth set describes also three fluids, but in this
case the parameter $\kappa_3$ is negative, leading
to a spontaneous encapsulation of liquid $C_2$ by liquid $C_3$.
Negative values of lambdas or kappas are generally allowed in the ternary model,
as long as the three minima in the $[\rho,\phi]$ space are well defined. 

\begin{figure*}[tbh]
\centering
\includegraphics[width=0.9\textwidth]{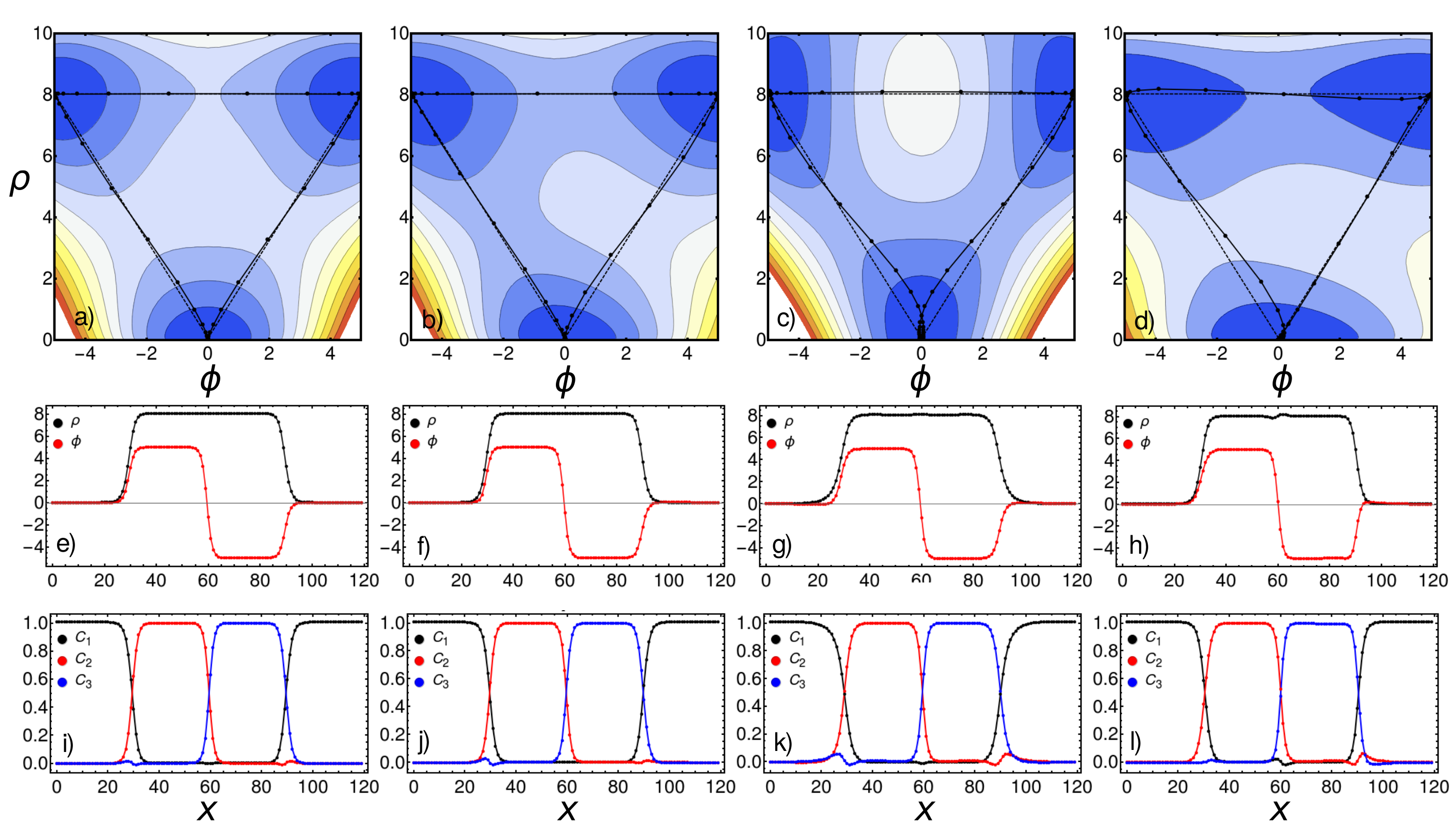}
\caption{\label{fig:freeenergy} 
(Color online) 
Upper row: color maps of the bulk free energy for the parameter sets 
1,2,3 and 4 (panels a,b,c and d). The dashed lines connecting the 
three energy minima represent the path in the of $[\phi,\rho]$  space
of an ideal interface, where the third components is completely absent. 
Data points represent the path of the numerically computed interface
profiles in mechanical equilibrium. Deviations from the dashed lines
reveal the creation of a fraction of concentration the third component. 
Middle row (panels e,f,g and h): Profiles of $\rho$ and $\phi$ along 
interfaces between fluids, placed in the sequence $1,2,3,1$.
Bottom row (panels i,j,k and l): Profiles of $C_1$, $C_2$ and $C_3$ 
along interfaces between fluids in the corresponding sequence $1,2,3,1$.
}
\end{figure*}

In Fig. \ref{fig:freeenergy} we inspect the properties of the diffuse 
interfaces for the parameter sets described in table \ref{tab:tabletension}. 
The color maps illustrate the contours of the bulk free energy in
the $[\phi,\rho]$ space. As expected, the bulk free energy is symmetric 
in $\phi$ for sets 1 and 3, and non-symmetric for sets 2 and 4. 

Introducing the variable transformation  Eqns. \ref{equ:definitionC1}, \ref{equ:definitionC2} 
and \ref{equ:definitionC3} into Eqn. \ref{equ:definitionCsum} we can 
easily see that the absence of the third component at any interface
leads to a linear relation between $\rho$ and $\phi$ connecting the
corresponding minima, represented by straight lines in the $[\rho,\phi]$ 
space in Fig. \ref{fig:freeenergy}.
However, the minimisation of the free energy does lead to different
paths, depicted by connected dots. As Eqn. \ref{equ:definitionCsum} 
must be satisfied, the inverse variable transformation will produce
a certain fraction of the minor component at the interface. 

\begin{figure}[b]
\centering
\includegraphics[width=0.5\columnwidth]{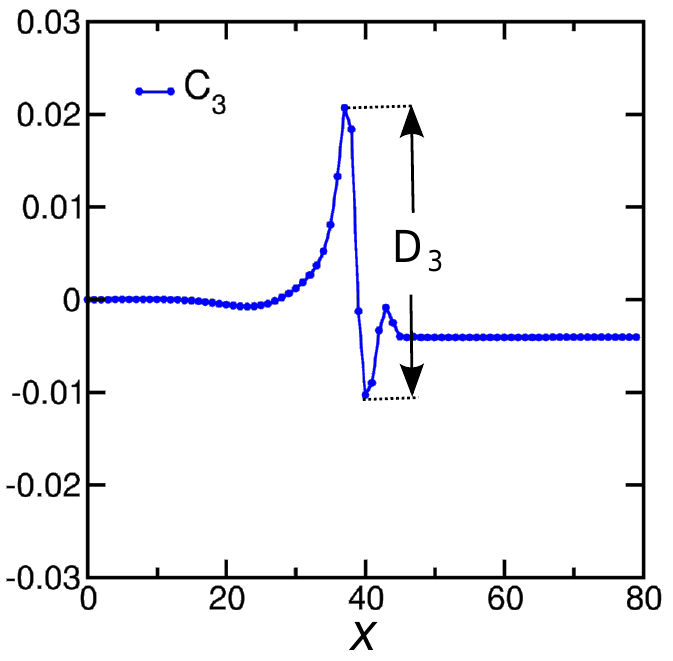}
\caption{\label{fig:deformationcoefficient}
(Color online) Example of profile of the concentration $C_3$ at the
interface between $C_1$ and $C_2$ illustrating the definition of the
Deformation Coefficient ($D_3$). 
}
\end{figure} 

For set 1 the interface path is close to the \enquote{ideal} profile. 
The deviations in the remaining sets increase with the increasing
mismatch between the profiles of $\rho$ and $\phi$.  
To quantify these mismatches we define the \enquote{Deformation coefficient} 
$D$ as the difference between the maximum and minimum values 
of the minority phase in a region $\Omega$ near the 
interface between the two majority phases.
For example, at the interface between $C_1$ and $C_2$, 
the Deformation coefficient $D_3$ is defined as
\begin{equation}
\label{equ:QFdef}
D_{3}=\max_{\vec{x}\in\Omega}(C_{3})-\min_{\vec{x}\in\Omega}(C_{3}),
\end{equation}	
and similarly for the other interfaces.
	
\begin{table}[bth]
\begin{tabular}{|c||c|c|c|c|c|c| }
\hline
{\bf subspace } & $\bm \lambda_1$ & $\bm \lambda_2$ & $\bm \lambda_3$ 
& $\bm \kappa_1$ & $\bm \kappa_2$ & $\bm \kappa_3$ \\
\hline
\hline
{\bf 1 } & $Y$ & $X$ & $X$ & $0.01$ & $X$ & $X$  \\ 
{\bf 2 } & $Y$ & $X$ & $X$ & $0.01$ & $0.5X$ & $0.5X$  \\ 
{\bf 3 } & $Y$ & $X$ & $0.5$ & $0.01$ & $X$ & $0.5$  \\ 
{\bf 4 } & $0.01$ & $X$ & $Y$ & $0.01$ & $X$ & $Y$  \\ 
{\bf 5 } & $0.5$ & $X$ & $Y$ & $0.01$ & $X$ & $Y$  \\ 
{\bf 6 } & $1.0$ & $X$ & $Y$ & $0.01$ & $X$ & $Y$  \\ 
{\bf 7 } & $Y$ & $X$ & $0.5$ & $0.01$ & $2X-0.5$ & $1-X$  \\ 
{\bf 8 } & $0.1$ & $X$ & $1.0$ & $0.01$ & $Y$ & $1+X-Y$  \\ 
\hline
\end{tabular}
\caption{\label{tab:parametersets}
 Summary of surface tension tests. Details for each case in supplementary information.
}
\end{table}

To quantify the interfacial properties arising in the ternary model,
we have carried out a systematic analysis for a wide range of parameters. 
Because a complete scan of the six-dimensional parameter space formed by
${\bm \lambda}=(\lambda_1,\lambda_2,\lambda_3)$ and
${\bm \kappa}=(\kappa_1,\kappa_2,\kappa_3)$
is too demanding, we have identified eight subspaces.
Each subspace is a two-dimensional map defined by the coordinates, $X$ and $Y$.
Subspaces are divided in a $20\times 20=400$ points grid, and for each
point, representing a specific parameter set, we perform three independent
measures of the surface tensions (one for each interface). 
Each surface tension is obtained from a bubble test, introducing 
a drop of radius $80$ lattice units, within a simulation domain of 
$320\times320$ lattice units. 
In all tests we set $T_{red}=0.61$, leading to an effective density ratio 
$\rho_l/\rho_g\simeq 10^3$.

\begin{figure*}[htb]
\centering
\includegraphics[width=0.8\textwidth]{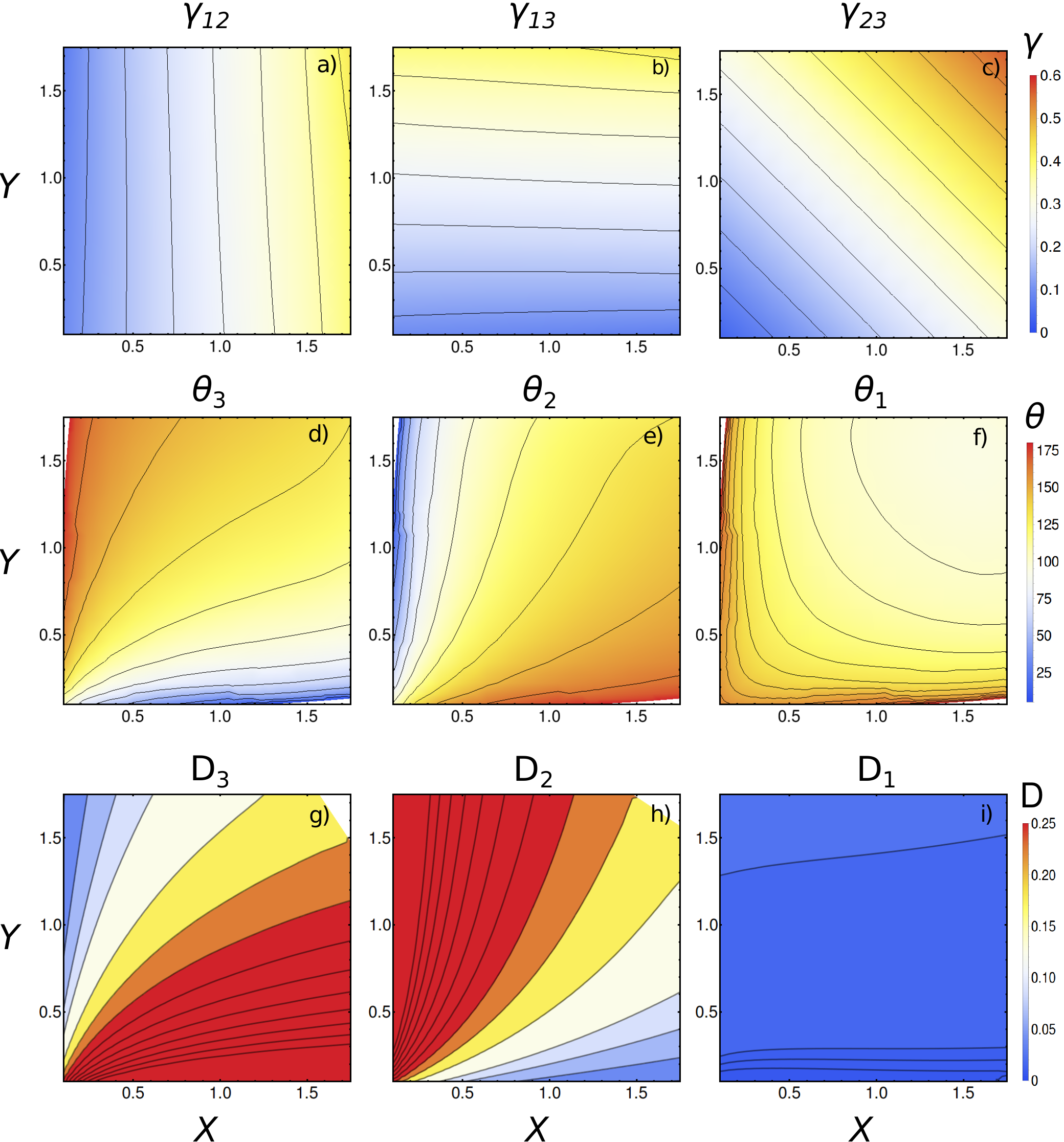}
\caption{\label{fig:tensionmap}
(Color online)
Colour maps of relevant quantities as function of the coordinate 
$X=\lambda_2$ and $X=\lambda_3$ 
Upper row (a,b,c): surface tensions ($\lambda_{12}$, $\lambda_{13}$ and $\lambda_{23}$); 
Middle row (d,e,f): Neumann angles ($\theta_{3}$, $\theta_{2}$ and $\theta_{1}$);
Lower row (g,h,i): Deformation coefficient ($D_{3}$, $D_{2}$ and $D_{1}$). 
}
\end{figure*}

The maps ${\bm \lambda}(X,Y)$ and ${\bm \kappa}(X,Y)$ 
are summarised in table \ref{tab:parametersets}.
In subspace 1 we consider liquids with identical properties 
$\lambda_2=\lambda_3=\kappa_2=\kappa_3=X$ and explore the 
relative impact of the liquid-gas component $\lambda_1=Y$.
Subspace 2 is similar, but the liquid-liquid width is larger
and the range of $\lambda_1=Y$ is extended to negative values, 
providing combinations of parameters for liquids repelling each other.
In subspace 3 we fix $\lambda_3=\kappa_3=0.5$, and explore the 
interplay between the gas phase and the first liquid. 
In subspace 4, 5 and 6 we fix the the weight of the 
bulk term for the equation of state to three values
$\lambda_1=0.01,0.6,1.0$, representing respectively small, medium and large 
contribution to the  liquid-gas component of the surface tension,
and systematically explore the combinations of the two liquids.
In subspaces 7 and 8 we explore combinations with negative 
values of $\kappa_2$, allowing us to achieve encapsulation
of liquid 3 by liquid 2. 

In Fig. \ref{fig:tensionmap} we report, as an example, our analysis
of the subspace 4. The first row of panels depicts the surface
tensions $\gamma_{12}$, $\gamma_{13}$ and $\gamma_{23}$ respectively. 
As expected, $\gamma_{12}$ and $\gamma_{13}$ mainly depend on the
variation of $X=\lambda_2$ and $Y=\lambda_3$ respectively, while 
$\gamma_{23}$ is function of $X+Y=\lambda_2+\lambda_3$. 
The non perfect alignment of the contour lines with the main axes
for $\gamma_{12}$ and $\gamma_{13}$ is an indication of the 
non constant contribution of the liquid-gas component, even if
$\lambda_1$ is fixed throughout the subspace. The variation of 
$\gamma_{23}$ instead is more regular, because no variation
of the density field occurs at this interface, and closely follows
the values of surface tension predicted by Eqn. \eqref{equ:tension23}
(comparison not shown).

The second row of panels in Fig. \ref{fig:tensionmap} reports the Neumann
angles $\theta_1$, $\theta_2$ and $\theta_3$ computed as functions of the
surface tensions:
\begin{equation}
\label{equ:theta1fromtension}
\cos\theta_1 = \frac{\gamma_{23}^2 - \gamma_{12}^2 - \gamma_{13}^2}{2 \gamma_{12}\gamma_{13}},
\end{equation}	
\begin{equation}
\label{equ:theta2fromtension}
\cos\theta_2 = \frac{\gamma_{13}^2 - \gamma_{12}^2 - \gamma_{23}^2}{2 \gamma_{12}\gamma_{23}},
\end{equation}	
\begin{equation}
\label{equ:theta3fromtension}
\cos\theta_3 = \frac{\gamma_{12}^2 - \gamma_{13}^2 - \gamma_{23}^2}{2 \gamma_{13}\gamma_{23}}. 
\end{equation}	
For the full range of parameters explored in this subspace, the
Neumann angles are always well defined, indicating that the spreading 
parameter $S_k=\gamma_{ij}-\gamma_{ik}-\gamma_{jk}<0$.
The third row of panels in Fig. \ref{fig:tensionmap} reports the 
\enquote{Deformation coefficient} $D$, measured for each interface. 
As expected, $D_1\simeq0$ throughout the whole map. 
In contrast, $D_2$ and $D_3$ vary up to $25\%$ of the concentration interval ($[0,1]$).

For each subspace in table \ref{tab:parametersets} we have fitted the maps 
of surface tension with a two-variable polynomial function. We report in the
supplementary material a detailed description, including fitting functions
for the surface tensions, of each subspace. This database can provide guidance 
to select the free energy parameters and reproduce target combinations 
of surface tensions.


\section{Solid Boundaries}
\label{boundaries}
		
In this section we describe and benchmark our implementation of solid boundaries. 
For simplicity we consider only flat walls, aligned with the domain 
axis and located at half distance between lattice nodes, but all
methods can be easily extended to solid structures with corners and wedges.
 
\begin{figure}[tb]
\centering
\includegraphics[width=0.8\columnwidth]{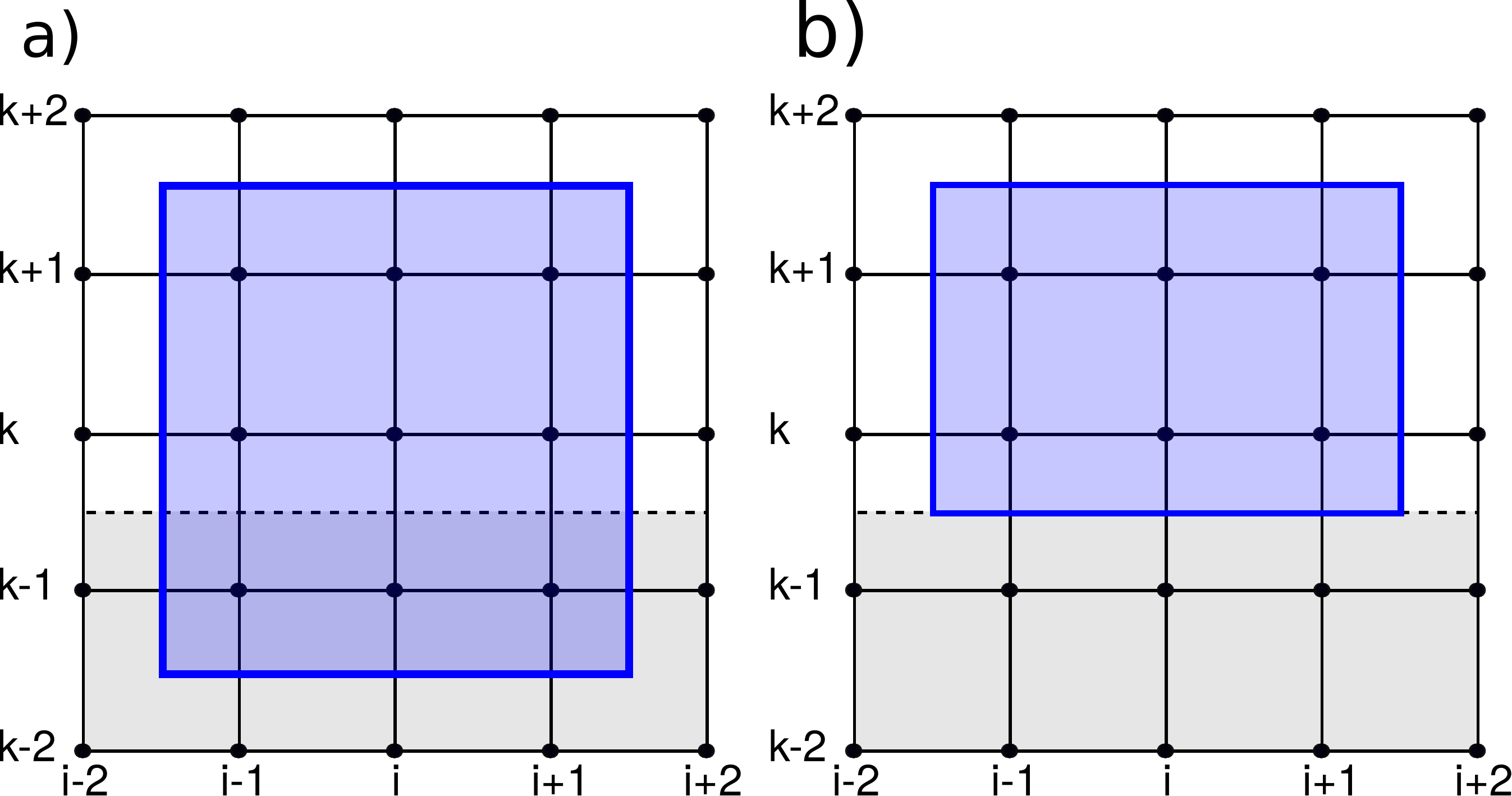}
\caption{\label{fig:gradient_stencils}
(Color online)
Sketch in 2D of the stencils employed for the computation of gradients.
(a) The stencil for $\nabla \rho$, $\nabla \phi$, $\nabla^2 \rho$ and
$\nabla^2 \phi$
is the same as in the fluid bulk, and relies on the quantities stored in the ghost
nodes in the solid layer.
(b) The stencil for $\vec{\nabla}\vec{P}_\parallel$ excludes solid nodes,
where $\vec{P}$ is not defined. }
\end{figure}

In all methods we treat the first layer of solid nodes as ghost 
nodes, to store values of $\rho$ and $\phi$. These values are 
employed in the finite difference stencils to compute 
$\nabla \rho$, $\nabla \phi$, $\nabla^2 \rho$ $\nabla^2 \phi$, 
in order to evaluate the chemical  potentials and the pressure tensor 
(Eqns. \eqref{equ:gradchempotrho}, \eqref{equ:gradchempotphi}, 
\eqref{equ:pressuretensor}) of the fluid near the solid boundaries,
as illustrated in Fig. \ref{fig:gradient_stencils} (a)

Throughout the whole fluid domain, the forces are computed by numerically 
differentiating the pressure tensor in Eqn. \eqref{equ:force}.  
As $\vec{P}$ is not defined on the solid nodes, its partial derivatives
in the first fluid layer are computed differently. Specifically, 
near the solid boundaries we impose $\vec{\nabla}\vec{P}_\perp=0$ (perpendicular to the solid), 
while a one-sided biased gradient \cite{Lee2010} is employed for the gradient 
$\vec{\nabla}\vec{P}_\parallel$ (parallel to the solid),
as illustrated in Fig. \ref{fig:gradient_stencils} (a). 
After the collision and streaming steps, standard bounce-back rules are 
applied \cite{Ladd1994}.

\subsection{Method 1 (force)}
\label{force}

\begin{figure}[tb]
\centering
\includegraphics[width=0.9\columnwidth]{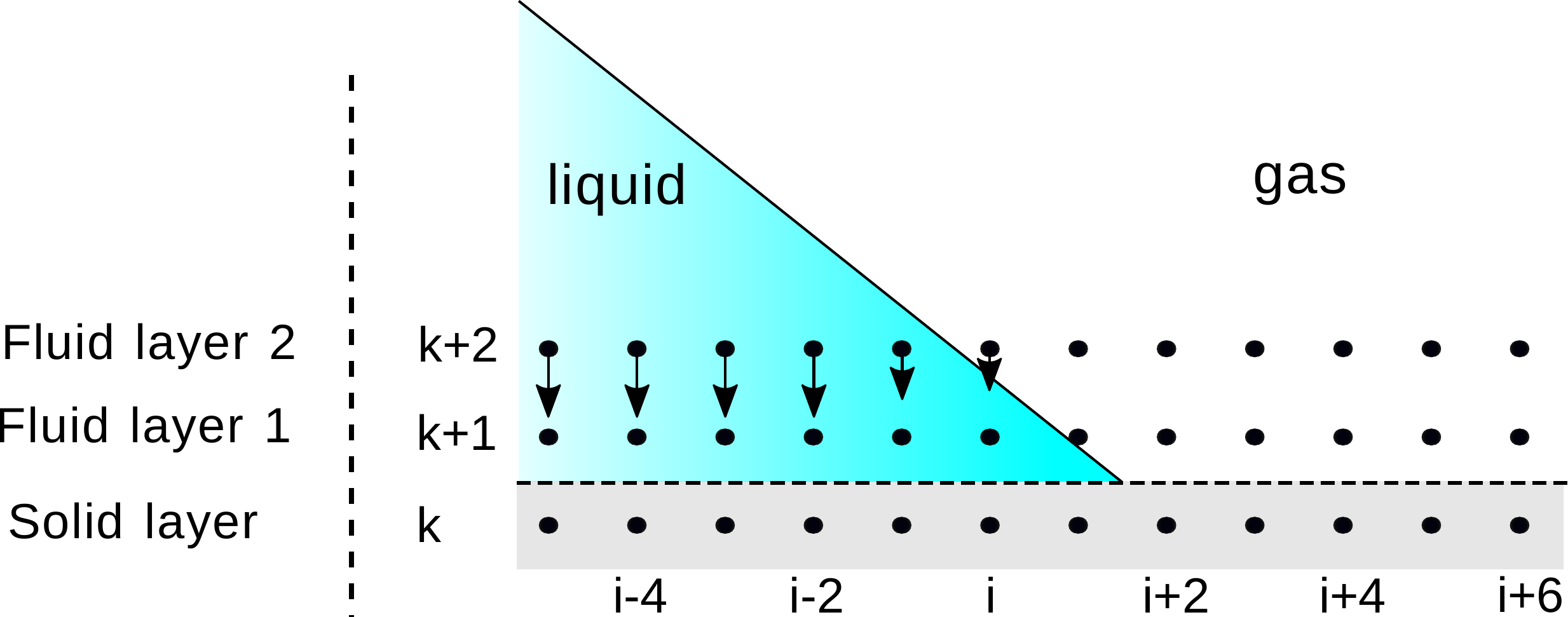}
\caption{\label{fig:force_Geom}
(Color online)
Sketch of the forcing terms acting near the liquid-liquid interface 
in contact to a solid boundary. The arrows represent the direction
and magnitude of the local force Eqn. \eqref{equ:forceBC}.}
\end{figure}

The forcing method \cite{Huang2007,M2015} is inspired by pseudo-potential 
models for multicomponent fluids, where the liquid-solid interaction
is introduced through a forcing term. 
In our implementation, the values of $\rho$ and $\phi$ in the
ghost nodes at the solid layer are constantly updated by
copying the values in the first fluid layer. This procedure 
alone gives to the solid neutral wetting properties. 
In this method, higher or lower affinity of the fluid phases 
to the solid are obtained by adding a local force term:
\begin{equation}
\label{equ:forceBC}
\vec{F}_s(\vec{x},t)=\rho^{rel}(\vec{x})\left(\kappa_\rho^w 
+ \phi^{rel}(\vec{x}) \kappa_\phi^w \right) 
\sum_i w_is(\vec{x}+\vec{c}_i\delta t)\vec{c}_i.
\end{equation}
where $s$ is a function that takes a value of $1$ on fluid nodes 
connected two lattice vectors away from solid nodes. In practice,
for a flat substrate as in the sketch in Fig. \ref{fig:force_Geom},
$s$ takes value 1 on the second layer of fluid nodes only.
We apply the force to the second fluid layer instead of 
the first one (as proposed in other works \cite{M2015}), 
to improve the stability of the algorithm. One can easily verify
that force terms of smaller magnitude are necessary at the second 
fluid layer to obtain the same target contact angle. 

The pre-factor $\rho^{rel}(\vec{x})\left(\kappa_\rho^w + 
\phi^{rel}(\vec{x}) \kappa_\phi^w \right)$
accounts for the variation of the interaction strength as
function of the fields $\rho$ and $\phi$, tuned by the 
parameters $\kappa_\rho^w $ and $\kappa_\phi^w $. 
We employ the rescaled fields $\rho^{rel}(\vec{x})=
(\rho(\vec{x})-\rho_g)/(\rho_l-\rho_g)$ 
and $\phi^{rel}(\vec{x})=\phi(\vec{x})/\chi$, which vary in 
the interval $[0,1]$ and $[-1.1]$ respectively. 

Furthermore, for the stability of the algorithm it is essential
that no large forcing terms are applied in the gas phase
($[\rho,\phi]=[\rho_g,0]$), which is achieved by multiplying 
both $\kappa_\rho^w $ and $\kappa_\phi^w $ by $\rho^{rel}(\vec{x})$
in our approach. In absence of this precaution, large forcing terms 
would cause strong deviations of the gas density from the 
equilibrium thermodynamic value.

When defining a contact angle between two phases we
indicate with the first index the phase in which
a contact angle is measured, and with the second index the other phase. 
In our work we adopt the convention of measuring the angles in the 
liquid phase at liquid-gas interfaces, while at the liquid-liquid
interface we measure the angle in the liquid with index 2. 
The following relations are implied: $\theta_{12}=\pi-\theta_{21}$, 
$\theta_{13}=\pi-\theta_{31}$ and $\theta_{32}=\pi-\theta_{23}$.

\begin{figure}[tb]
\centering
\includegraphics[width=0.98\columnwidth]{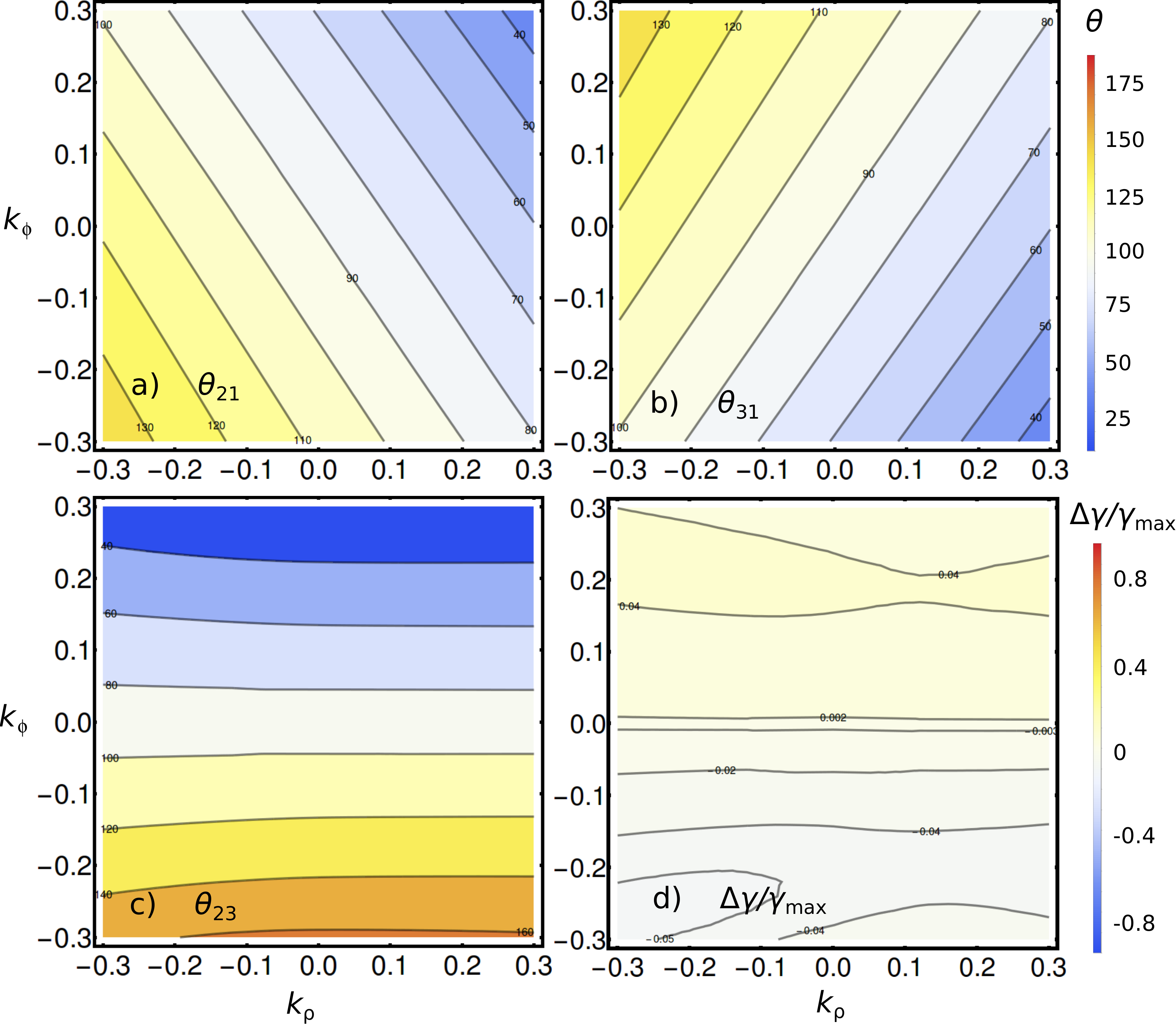}
\caption{\label{fig:force_angles}
(Color online)
Colour maps of equilibrium contact angles measured from sessile drops, 
as function of $\kappa_\rho^w$ and $\kappa_\phi^w$.  Panels a-c) refer
to the interfaces $[1,2]$,  $[1,3]$ and $[2,3]$ respectively. 
Panel d) reports the quantity $\Delta\gamma$, (Eqn. \ref{equ:GirifalcoGood}).
The combinations of surface tensions are given by the first 
set in table \ref{tab:tabletension}.
}
\end{figure}

A typical dependence of  $\theta_{21}$, $\theta_{31}$ 
and $\theta_{23}$ from the parameters $\kappa_\rho^w$ and $\kappa_\phi^w$ 
is reported in Fig. \ref{fig:force_angles} (panels a-c), for the
parameter set 1 in table \ref{tab:tabletension}.
Contact angles are measured after equilibrating 2D sessile drops 
for each interface and fitting the drop interfaces with circular profiles. 
To keep the accuracy of the contact angle, across the whole parameter range, 
the drop area is fixed to $\simeq 100^2$l.u.$^2$ while the size 
and aspect ratio of the simulation domain is adjusted to 
accommodate drops of different shapes. 

The maps, as shown in Fig. \ref{fig:force_angles}, are specific for each 
set of free energy parameters. For example
the inclination of the diagonal contour lines of $\theta_{21}$, $\theta_{31}$ 
depends on the value of the surface tension for each interface. $\theta_{23}$ 
instead is predominantly a function of $\kappa_\phi^w$, with only a residual 
dependence on $\kappa_\rho^w$ in the region of small $\kappa_\phi^w$.

On ideal surfaces the combinations of contact angles are not
independent, but obey the Girifalco-Good relation \cite{Girifalco1957},
which according to our convention reads
\begin{equation}
\label{equ:GirifalcoGood}
0=\Delta\gamma=\gamma_{23}\cos\theta_{23}-\gamma_{13}\cos\theta_{31}+\gamma_{12}\cos\theta_{21}
\end{equation}	
This condition is automatically satisfied by the force method, as can be deduced from
panel d) in Fig. \ref{fig:force_angles}, which reports the variation of $\Delta\gamma$ 
on a scale set by the largest value surface tension in the system ($\gamma_{\rm max}\simeq0.5$).
Small deviations identified by the contour lines can be attributed to 
the uncertainties in the measurement of the contact angles.

\subsection{Method 2 and 3 (geometric approaches)}
\label{geometrygradient}

We now introduce the two geometric approaches employed in our model. 
The key idea in both models is to manipulate the values of the fields 
in the ghost nodes at the solid boundaries according to a geometrical
criterion, in order to reproduce a prescribed contact angle.
In both cases, the ternary implementation requires us to identify 
in advance the correct interface, in order to select the correct
target contact angle. This step is performed by implementing a set of
rules that combine the local value of $\rho$ and $\phi$ and of their
gradients parallel to the solid $\nabla_\parallel\rho$ and $\nabla_\parallel\phi$: 
\begin{equation*}
\begin{cases}
\text{if } |\nabla_\parallel \rho| /|\nabla_\parallel \phi|<0.01(\rho_l-\rho_g)/\chi &\text{set interface 2-3}\\
\text{if } \nabla_\parallel \rho \cdot \nabla_\parallel \phi<0 &\text{set interface 1-3} \\
\text{if } \nabla_\parallel \rho \cdot \nabla_\parallel \phi>0 &\text{set interface 1-2} \\
\text{if } \rho>\rho_l/2 \text{ and } \phi<-0.95\chi &\text{set interface 1-3} \\
\text{if } \rho>\rho_l/2 \text{ and } \phi>0.95\chi &\text{set interface 1-2} \\
\end{cases}
\end{equation*}
This set of rules proves to be accurate in all our tests, even if the
variation of $\rho$ and $\phi$ is not strictly monotonous near the interface.
An alternative approach consists in weighting the contact angles based
on the local concentration fields \cite{Zhang2016a}.

\begin{figure}[bt]
\centering
\includegraphics[width=0.6\columnwidth]{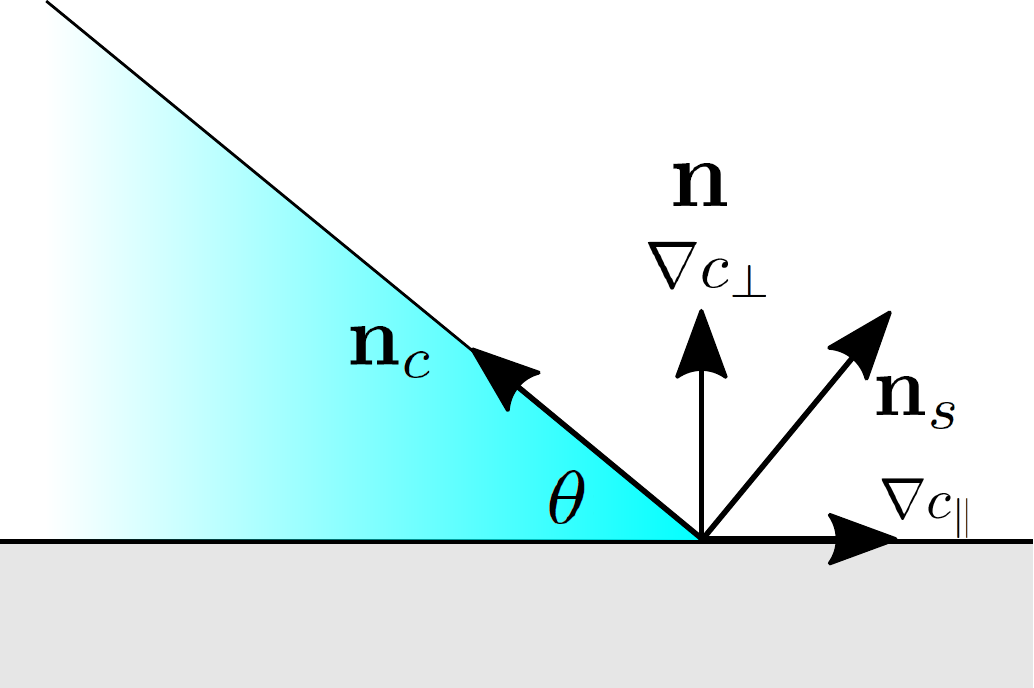}	
\caption{\label{fig:geometricextrapolation}
(Color online)
Sketch of the main vectors defined by fluid interface near the contact line.
}
\end{figure}

\emph{Geometric extrapolation -}
We now introduce our ternary implementation of the method proposed 
by Ding and Spelt \cite{Ding2007}. The key idea is to compute the normal 
vector of a fluid interface in contact with the solid surface from 
the gradient of a field: $\vec{n}_s=\nabla c/|\nabla c|$.
We employ $c= \rho$ at any liquid-gas interface, and $c= \phi$ for
the liquid-liquid interface. Referring to the sketch of the contact line
geometry in Fig. \ref{fig:geometricextrapolation}, $\vec{n}$ defines the 
vector normal to a plane solid surface, while the perpendicular and parallel 
components of a field gradient can be expressed as $\nabla c_\perp=\vec{n}\cdot\nabla c$ 
and $\nabla c_\parallel=|\nabla c -(\vec{n}\cdot\nabla c) \vec{n}|$. 

In the algorithm, first the parallel component of the gradient $\nabla c_\parallel$ 
is measured along the surface, and then it is employed to reconstruct the 
perpendicular component of the gradient $\nabla c_\perp$.
For a diffuse interface forming an angle $\theta$ with the solid surface, 
the relation between components of the field gradients is 
\begin{equation}
\label{equ:geomamgle}
\nabla c_\perp=\tan\left(\frac{\pi}{2}-\theta\right)\nabla c_\parallel.
\end{equation}

For example, in a 2D lattice addressed by the indices $i,k$, let us assume
the layer $k$ represents a solid surface for any $i$, while the
layer $k+1$ represents the first fluid layer. The values of the
field $c_{i,k}$ are computed by extrapolating the from the field value 
in the above layer 
\begin{equation}
\label{equ:geomding}
c_{i,k}=c_{i,k+1}+\nabla c_\perp,
\end{equation}
where $c$ represents either $\rho$ or $\phi$. For this reason, we denote this
method as \emph{geometric extrapolation}.

The 3D implementation differs from the 2D case only by replacing the component  
parallel to the surface of the concentration gradients with the norm of the two 
components in the plane. For example, if $x$ and $y$ define the coordinates
in the plane, we have $|\nabla_\parallel c|=\sqrt{(\nabla_x c)^2+(\nabla_y c)^2}$
for a solid plane at $z={\rm const}$. The correction applies both in the determination 
of the interface, and in the reconstruction of the perpendicular component $\nabla_\perp c$.

In contrast to the force approach, there is no limitation in choosing 
any combination of contact angles, keeping in mind that a physically consistent 
set of contact angles needs to fulfil the Girifalco-Good relation in Eqn. \eqref{equ:GirifalcoGood}.
We will take advantage of combinations of angles not fulfilling Eqn. \eqref{equ:GirifalcoGood}
to simulate self propelled bi-slugs in a channel in Sec. \ref{bislug}.

\emph{Geometric interpolation -}
This third method is inspired by the algorithm proposed 
by Lee and Kim \cite{Lee2011}. Here the key idea is 
to interpolate the field values from the upper layer, where
the interpolating point is shifted according to the slope
of the liquid interface.  

For a few special values of contact angles the slope of the interface 
connects to lattice nodes, and the required values of the field 
correspond exactly to the values already stored. 
Let us consider a 2D example: for the three nearest lattice nodes 
along the direction $i$ of the solid surface we can simply assign
\begin{align}
c_{i,k}|_{\theta\simeq 18.43^\circ}&=c_{i-3,k+1}   \nonumber \\
c_{i,k}|_{\theta\simeq 26.56^\circ}&=c_{i-2,k+1}  \nonumber\\
c_{i,k}|_{\theta=45^\circ}&=c_{i-1,k+1}   \nonumber\\
c_{i,k}|_{\theta=90^\circ}&=c_{i,k+1} \nonumber \\
c_{i,k}|_{\theta=45^\circ}&=c_{i+1,k+1}   \nonumber\\
c_{i,k}|_{\theta\simeq 153.43^\circ}&=c_{i+2,k+1}  \nonumber\\
c_{i,k}|_{\theta\simeq 161.56^\circ}&=c_{i+3,k+1}.   \nonumber \\
\end{align}

\begin{figure}[tb]
\centering
\includegraphics[width=0.98\columnwidth]{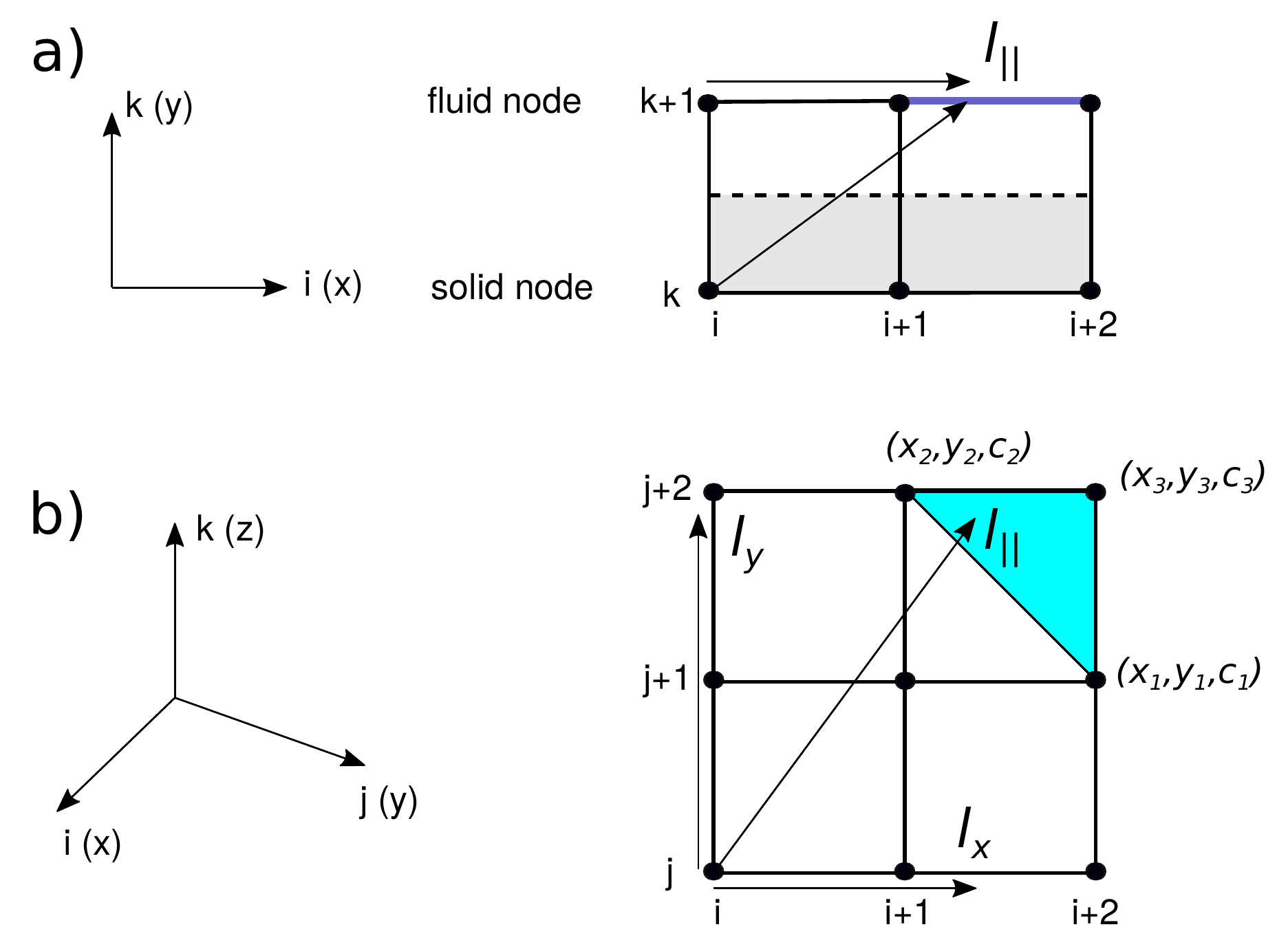}	
\caption{
\label{fig:geometricinterpolation}
(Color online)
Sketch of the \emph{geometry interpolation} boundaries: 
a)2D implementation: the selected interval for the linear interpolation is highlighted.
b)3D implementation: the selected triangle for the planar interpolation is highlighted. 
}
\end{figure}

For any other slope instead we linearly interpolate the values of the two 
closest nodes. For this reason  we denote this method as \emph{geometric interpolation}.
As shown in Fig. \ref{fig:geometricinterpolation} (a), in the 2D
implementation we compute the distance of the interpolating point from the node $i$ 
as $l_\parallel=\tan(\theta-\pi/2)$. In a local coordinate system centred in 
the node $(i,k)$, the interpolating points are located at $l_0={\rm floor}(l_\parallel)$ 
and $l_1={\rm floor}(l_\parallel)+1$, and their lattice indices are $i_0=i+l_0$ and $i_1=i+l_1$.
Considering that $l_1-l_0=1$, the linear interpolation scheme is
\begin{equation}
\label{equ:interp2D}
c_{i,k}=(c_{i_1,k+1}-c_{i_0,k+1})l_\parallel+(l_1 c_{i_0,k+1} - l_0 c_{i_1,k+1}).
\end{equation}

The 3D implementation requires the selection of an appropriate support for 
the interpolation in the plane. As in the previous case, let us assume a solid 
surface defined by the plane plane $z={\rm const}$, where solid nodes have constant
index $k$ and the first fluid layer is at $k+1$. Also, the lattice nodes
in the planes parallel to the solid surface are addressed by the indices $i$ and $j$.
The location of the interpolating points is determined by the gradients of the concentration
field in the plane $l_x=l_\parallel \nabla_x c/|\nabla c_\parallel|$ and 
$l_y=l_\parallel \nabla_y c/|\nabla c_\parallel|$. The simplest interpolation scheme for
a plane in 3D requires three points. In our implementation we select the three furthest
points (out of four) from the location $(i,j)$ in a planar square lattice
(cfr. the sketch in Fig. \ref{fig:geometricinterpolation} (b). 

Once the three points are selected, we consider the three triplets 
$(x_1,y_1,c_1)$, $(x_2,y_2,c_2)$ and $(x_3,y_3,c_3)$,
describing a plane, where the third coordinate 
represents the value of the concentration $c$ in each point.
The following interpolation scheme is employed to compute 
the field values in the ghost node located in $(i,j,k)$:
\begin{equation}
\label{equ:interp3D}
	c_{i,j,k}=\frac{A l_x+ B l_y + C}{D},
\end{equation}
where $A,B,C,D$ are the polynomials
\begin{align}
	A =&y_2 c_1 - y_3 c_1 - y_1 c_2 + y_3 c_2 + y_1 c_3 - y_2 c_3 \nonumber \\
	B =&-x_2 c_1 + x_3 c_1 + x_1 c_2 - x_3 c_2 - x_1 c_3 + x_2 c_3  \nonumber\\
	C =&-x_3 y_2 c_1 + x_2 y_3 c_1 + x_3 y_1 c_2  \nonumber\\
	& -x_1 y_3 c_2 - x_2 y_1 c_3 + x_1 y_2 c_3  \nonumber\\
	D =&- x_2 y_1 + x_3 y_1 + x_1 y_2 - x_3 y_2 - x_1 y_3 + x_2 y_3.
\end{align}

\begin{figure}[t]
\centering
\includegraphics[width=0.98\columnwidth]{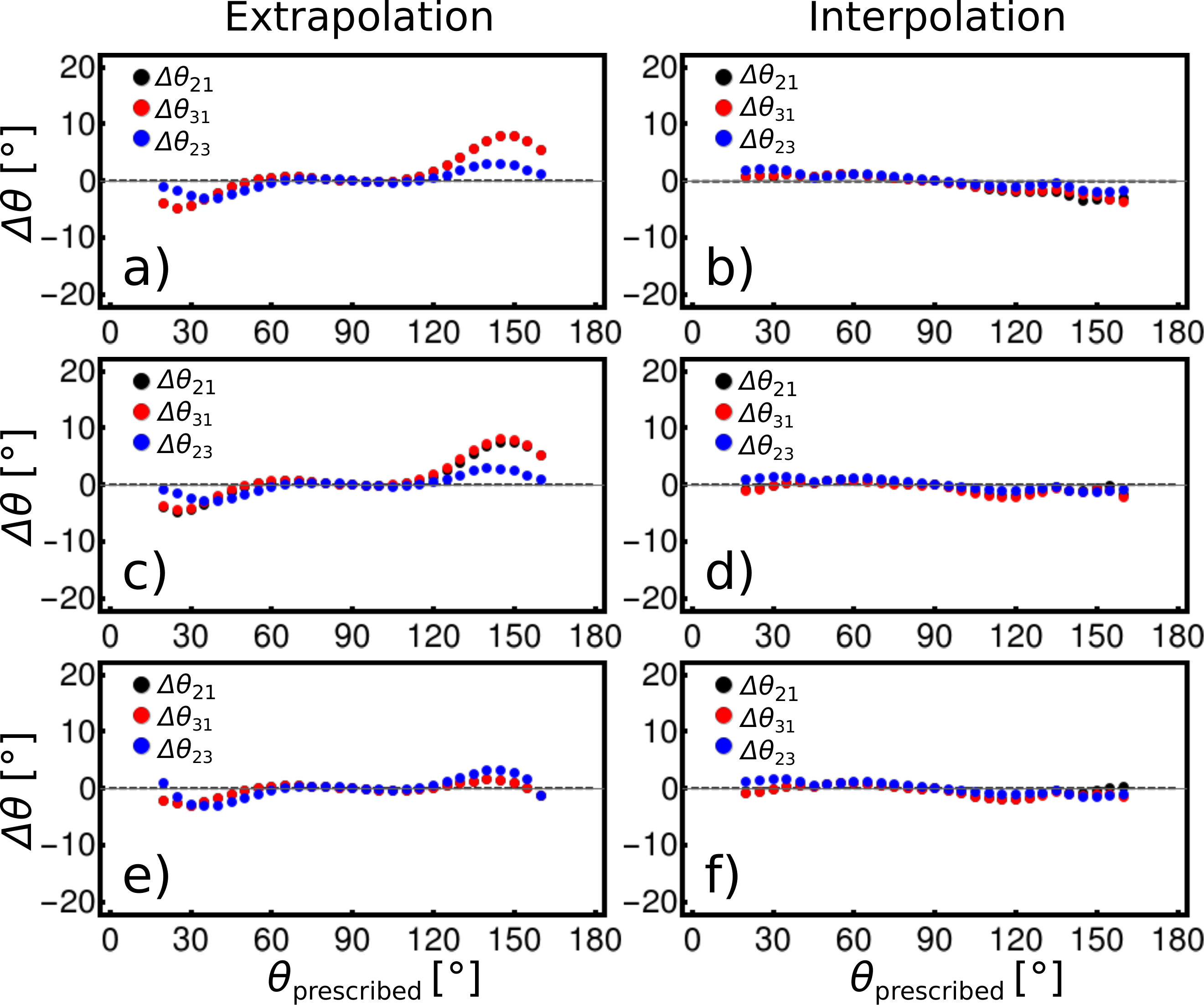}
\caption{\label{fig:geom_angles} 
(Color online)
Deviation of contact angles $\theta_{21}$, $\theta_{31}$ and $\theta_{23}$,
measured on sessile drops in mechanical equilibrium from the prescribed values.
Results for the  \emph{geometric extrapolation} method (left column)
and the \emph{geometric interpolation} method (right column). 
The interfacial properties correspond to set 1 (first row); 
set 2 (second row); and set 3 (first row) listed in table \ref{tab:tabletension}. 
}
\end{figure}
			
We have assessed the accuracy of both geometric methods by simulating 
sessile drops in mechanical equilibrium for each fluid-fluid interface 
and comparing the parameter sets 1, 2 and 3 in table \ref{tab:tabletension}.
The simulation setup and analysis are the same as previously employed to validate 
the force method. In the intermediate range of angles $[60^\circ,120^\circ]$ 
both methods show good agreement, with deviations below $1^\circ$, while 
for larger and smaller angles the \emph{geometry interpolation} method
is to be preferred.
In view of this result, we discard the \emph{extrapolation}
in favour of the \emph{interpolation} method for the remaining tests.

\section{Capillary filling}
\label{filling}

To assess the dynamic properties of fluid interfaces, 
we simulate the capillary filling of a channel by a liquid. 
The problem was studied independently by Richard Lucas \cite{Lucas1918} and Edward 
Washburn \cite{Washburn1921}. It represents a classical benchmark for wetting boundary 
conditions in lattice Boltzmann implementations \cite{Wiklund2011,Liu2015d,Lou2018}, 
as it provides analytical or semi-analytical expressions to compare.

\begin{figure}[tb]
\centering
\includegraphics[width=0.9\columnwidth]{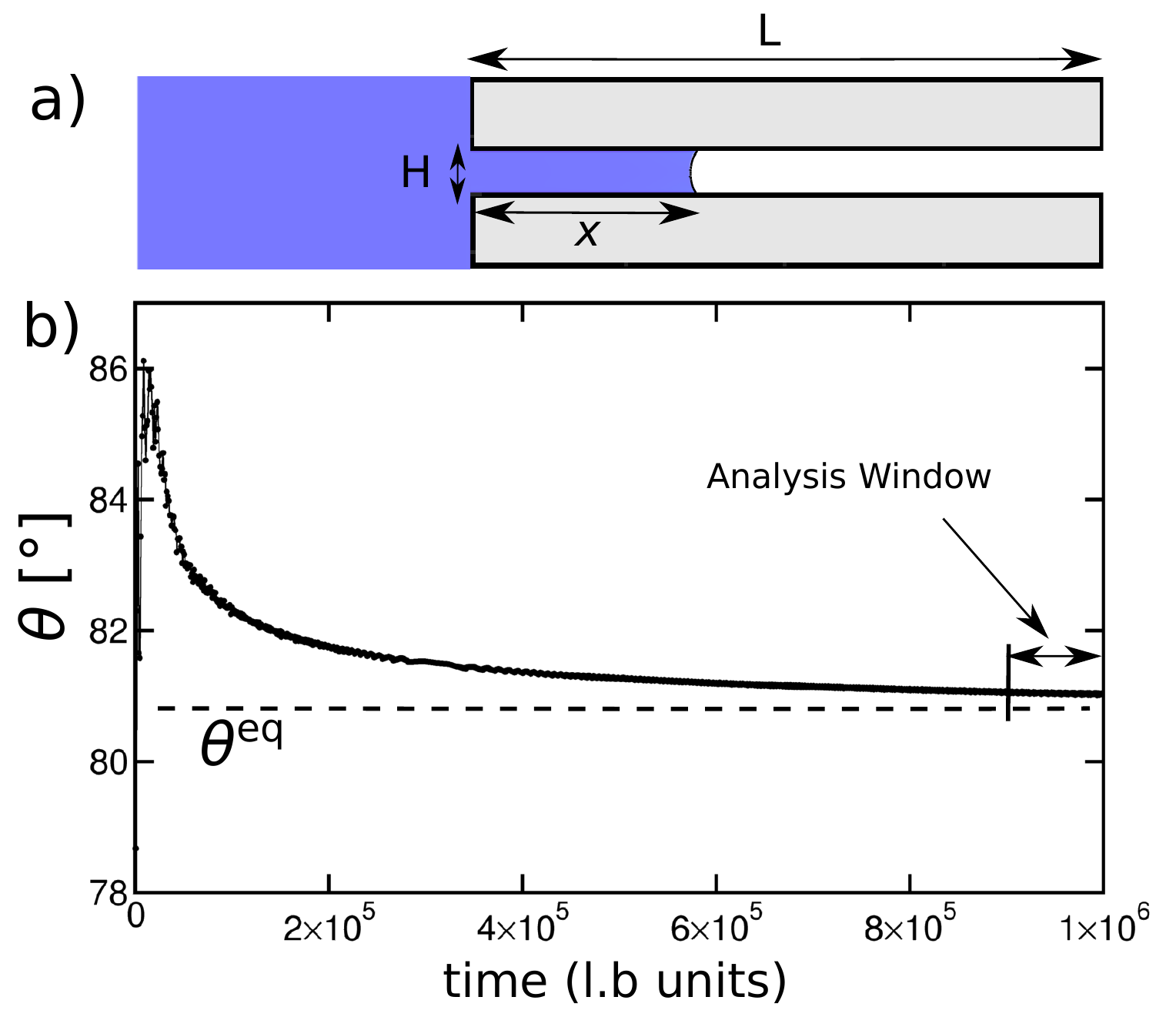}
\caption{\label{fig:capillary_filling_sketch} 
(Color online) 
Capillary filling:
(a) Sketch of the simulation setup; 
(b) Dynamic contact angle vs simulation time for the \emph{Geometric Interpolation} 
method and $\theta_{21}=80^\circ$.
The analysis window consists of the last $10\%$ of the simulation time, 
over which the dynamic contact angle is averaged.
 }
\end{figure}

Let us now consider the system sketched in Fig. \ref{fig:capillary_filling_sketch} ,
consisting in a 2D channel of height $H$, initially containing a gas phase only,
and filled by liquid. The liquid-gas surface tension is denoted by $\gamma$, 
while the liquid forms a contact angle $\theta$ with the solid.

In a 2D geometry, the driving capillary force is applied at the two 
contact points of the liquid interface with the channel walls:	
\begin{equation}
\label{equ:capforce}
F^{\rm cap}=2\gamma\cos\theta.
\end{equation}
Except for the initial transient time, the resisting force is mainly provided
by viscous dissipation. In virtue of the high density ratio
in our model, we neglect the dissipation in the gas phase \cite{Diotallevi2009b}.
For a liquid of viscosity $\mu=\rho_l \nu$ forming a column of length $x$, 
and assuming a fully developed Poiseuille velocity profile, we have
a resisting force 
\begin{equation}
\label{equ:viscousforce}
F^{\rm visc}=-\frac{12 \rho_l \nu x \dot{x}}{H},
\end{equation}
where $\dot{x}$ is the mean velocity of the fluid column, corresponding to the 
velocity of the liquid-gas interface.
Eqns. \eqref{equ:capforce} and \eqref{equ:viscousforce} lead to the 
so-called Washburn law
\begin{equation}
\label{equ:Washburn}
x(t)=K\sqrt{(t-t_0)},
\end{equation}
where $t_0$ is a time constant. The pre-factor 
\begin{equation}
\label{equ:Washburnprefactor}
K=\sqrt{\frac{\gamma\cos\theta H}{3\nu\rho_l}},
\end{equation}
is function of material and geometric parameters only: the surface tension $\gamma$,
the equilibrium contact angle $\theta$, the kinematic viscosity $\nu$
of the liquid and channel height $H$.

\begin{figure}[tb]
\centering
\includegraphics[width=0.9\columnwidth]{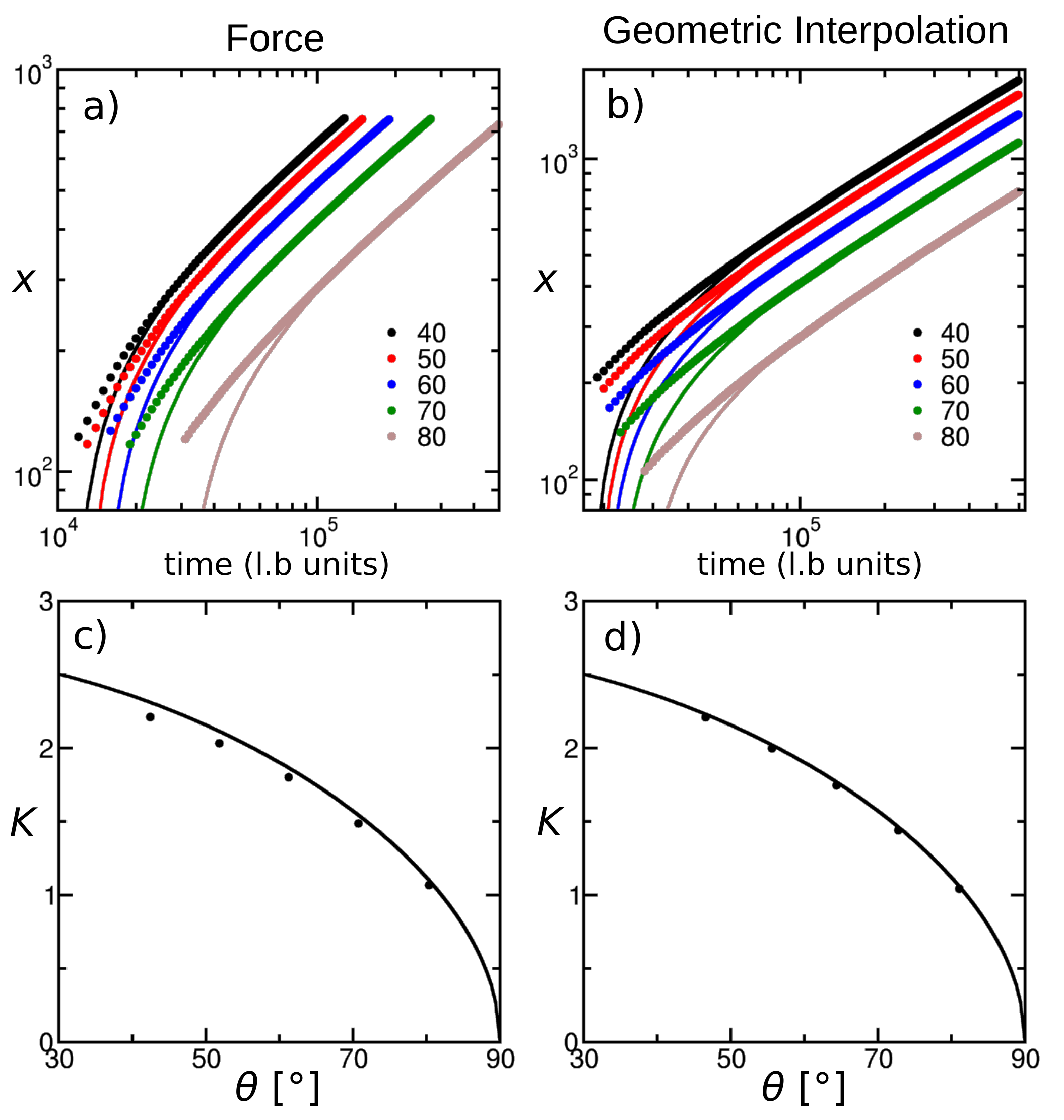}
\caption{\label{fig:capillary_filling} 
(Color online) 
Capillary filling:
(a,b) Length of the liquid column vs simulation time for 
contact angles $\theta=40^\circ,50^\circ,60^\circ,70^\circ$ and $80^\circ$.
Dots represent numerical results, while continuous lines data fits.
(c,d) Pre-factor $K$ for the Lucas-Washburn law. Dots represent
fits to the numerical results, the continuous line is the 
model prediction (Eqn. \eqref{equ:Washburnprefactor}).
Left panels (b,d) are obtained employing the \emph{force} method,
while right panels (c,e) employ the \emph{geometric interpolation} 
method.   
 }
\end{figure}

In our simulations the channel length is $L=2000$ l.u. and the height $H=70$l.u.  
The channel is preceded and followed by reservoirs filled with liquid and gas respectively. 
This geometry has been previously employed \cite{Diotallevi2009b} to minimise 
the viscous drag of the fluid outside the channel. Throughout all simulations
we employ the first parameter set in table \ref{tab:tabletension}, for which
$\gamma=0.414$ in both liquid-gas interfaces, and set $\beta=0.5$, giving a 
kinematic viscosity $\nu=1/6=0.16667$. 

In Fig. \ref{fig:capillary_filling} (a,b) we report the time evolution
of the front of the liquid column for contact angles varied in the range of
$[30^\circ,80^\circ]$, comparing the \emph{force} and 
\emph{geometric interpolation} methods.
The initial stage of the invasion is not well described 
by Wahsburn law \cite{Diotallevi2009b}. 
As shown in Fig.  \ref{fig:capillary_filling_sketch} (b), during the 
filling process the dynamic angle varies over time, and approaches 
the equilibrium value only asymptotically. Consequently,
Washburn law, Eqn. \ref{equ:Washburn} describes accurately only the
asymptotic regime, while for the initial and transient regimes
inertia and viscous bending should also be considered.

As in this specific test our main interest is comparing the accuracy
of the \emph{force} and \emph{geometric interpolation} methods,
we analyse the last $10\%$ of the simulation time, where the
variation of dynamic angles are below one degree, and we can
assume Eqn. \ref{equ:Washburn} to be sufficiently accurate.
To eliminate systematic sources of errors, we compute the
average dynamic angle  $<\theta>$  within the analysis window,
and replace with it the angle $\theta$ in Eqn. \eqref{equ:Washburnprefactor}.
Furthermore, we perform a parametric fit of the numerical data 
within the analysis window with Eqn. \eqref{equ:Washburn}, 
and compute the time constant $t_0$ and the pre-factor $K$.
 
In Fig. \ref{fig:capillary_filling} (c,d) we compare the values of $K$ 
to the model (Eqn. \eqref{equ:Washburnprefactor}).
The data for the \emph{force} method show small deviations between
predicted and measured values of the prefactor. The deviations 
increase proportionally with the magnitude of the forcing term,
which increases as $\theta$ decreases. 
This suggests that the discrepancy is related to spurious velocities
near the walls, due the force term in the \emph{force} method.
In contrast, we observe no deviations for the \emph{geometric interpolation}.


\section{Self-propelled slugs}
\label{bislug}

In this section  we focus on a ternary-specific benchmark, consisting in a 
self-propelled train of drops (bi-slug) in a 2D channel.

In experiments, a bi-slug with three finite contact angles can not self-propel, 
unless the Girifalco-Good relation, Eqn. \eqref{equ:GirifalcoGood}, is broken.
This may be done by introducing a step or gradient of wettability on the channel 
surfaces\cite{Esmaili2012,Huang2014}. 
Alternatively, at least one liquid phase must be completely wetting.
This last condition was exploited by Bico and Quer\'e to study experimentally
in detail self propelled bi-slugs \cite{Bico2000,BICO2002a}. 
Taking advantage of the \emph{geometric interpolation} method, we can numerically 
introduce arbitrary contact angles in the system providing a controlled mechanism
for self-propulsion. 

The simulation geometry, sketched in Fig. \ref{fig:Bislugfig} (a), consists of
a periodic channel of height $H=39$ l.u. It contains a train of drops 
having equal volumes. For simplicity, we assume the length $L_1=L_2$ of each 
liquid drop, approximated by the length of the equivalent rectangle
having the same area and height $H$. The total length of the periodic channel
is adjusted in each simulations to allow the presence of at least $200$ lattice 
units of gas at the two sides of the bi-slug.

\subsection{Bi-slug dynamics}

In long trains of drops the driving force is almost completely
dissipated in the liquid bulk. Consequently the velocity is small and the 
contact angles remain close to the equilibrium value. 
According to the convention for contact angles employed in this work, 
the surface tension unbalance is expressed by 
\begin{equation}
\label{equ:capillaryforce}
\Delta\gamma=\gamma_{23}\cos\theta_{23}-\gamma_{13}\cos\theta_{31}+\gamma_{12}\cos\theta_{21},
\end{equation}
and the driving force is 
\begin{equation}
\label{equ:capillarydrive}
F^{\rm cap}=2\Delta\gamma,
\end{equation}
Assuming  a Poiseuille flow profile in the bi-slug, 
and liquids with equal viscosity, the viscous force is
\begin{equation}
\label{equ:viscousfriction}
F^{\rm visc}=-\frac{12 \rho_{l}\nu L \dot{x}}{H},
\end{equation}
where $\dot{x}$ is the mean fluid velocity, associated
to the velocity of the center of mass of the bi-slug.

In the limit of long trains ($L=L_1+L_2>>H$) the viscous bending
can be neglected, and the equation of motion for the center of mass 
is \cite{Diotallevi2009b,Esmaili2012}
\begin{equation}
\label{equ:bislugmotion}
\rho_l LH \ddot{x}=F^{\rm cap}+F^{\rm visc}.
\end{equation}
By introducing Eqns. \eqref{equ:capillarydrive} and \eqref{equ:viscousfriction}
into Eqn. \eqref{equ:bislugmotion}, we obtain
\begin{equation}
\label{equ:bislugdifferential}
\ddot{x} +\frac{12\nu}{H^2}\dot{x}-\frac{2\Delta\gamma}{\rho_l LH}=0.
\end{equation}
Integrating once with time and imposing $\dot{x}(0)=0$, 
we obtain an exponential relaxation of the bi-slug velocity 
to the steady velocity $v_{\infty}$
\begin{equation}
\label{equ:centroidvelocity}
\dot{x}(t)=v_{\infty}(1-e^\frac{-t}{\tau_{\rm rel}}),
\end{equation}
where  the steady velocity is
\begin{equation}
\label{equ:bislugvelocity}
v_{\infty}= \frac{H \Delta \gamma}{6 \rho_{l} \nu L}
\end{equation}
and the relaxation time is
\begin{equation}
\label{equ:bislugrelaxation}
\tau_{\rm rel}= \frac{H^2 }{12 \nu}.
\end{equation}
Integrating Eqn. \eqref{equ:centroidvelocity} once again with time we obtain
the displacement of the center of mass with respect to its initial position $x(0)$
\begin{equation}
\label{equ:centreofmassposition}
x(t)=x(0)+\tau_{\rm rel} v_{\infty}(e^\frac{-t}{\tau_{\rm rel}}+\frac{t}{\tau_{\rm rel}}-1).
\end{equation}

\begin{figure}[tb]
\centering
\includegraphics[width=0.98\columnwidth]{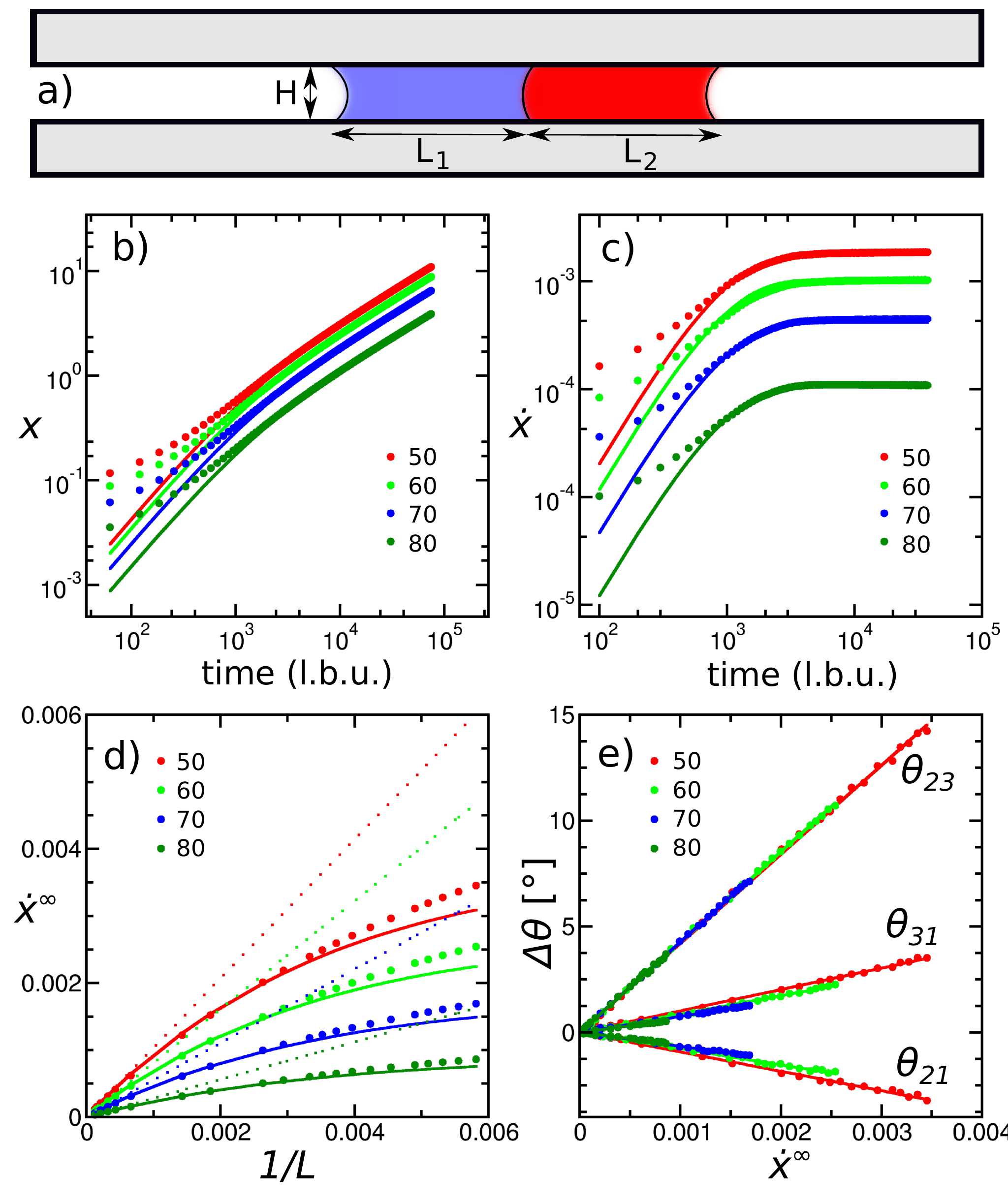}
\caption{
\label{fig:Bislugfig} 
(Color online) Self- propeled bi-slugs:
(a) Sketch of the simulation setup.
The other panels report data for bi-slugs defined by the first parameter set 
in table \ref{tab:tabletension}. We set $\beta=0.5$ and contact angles 
$\theta_{23}$=$\theta_{21}$=$\theta_{31}=50^\circ,60^\circ,70^\circ,80^\circ$.
(b,c) Transient regime in the motion of the bi-slugs of length $L=1500$l.u.,
showing (b) the position and (c) the velocity of the center of mass.
(d) The measured steady velocity $\dot{x}_\infty$ as function of $1/L$.
(e) The dynamic contact angles for the three interfaces as function of $\dot{x}_\infty$. 
}	
\end{figure}

In the simulation we initialise the bi-slug as two rectangular liquid drops
in the beginning of the channel. Typically, during the first $10^3$ steps 
of the simulation, the liquid interfaces quickly deform to approach the contact 
angle of the steady moving bi-slug, and initiate the self-propulsion mechanism.
In Fig. \ref{fig:Bislugfig} we compare the trajectory (panel (b)) 
and velocity (panel (c)) of a long train of drops ($L=1500$l.u.),
with Eqns. \eqref{equ:centreofmassposition} and \eqref{equ:centroidvelocity},
for contact angles $\theta_{23}$=$\theta_{21}$=$\theta_{31}$ varied 
in the range $[50^\circ,80^\circ]$.
Eqns. \eqref{equ:centroidvelocity} and \eqref{equ:centreofmassposition} 
capture accurately the bi-slug dynamics after the first $10^3$ 
time steps. A close inspection of panel (c) shows that 
after $10^4$ time steps the bi-slug speed has fully 
reached the steady value $v_\infty$. 

In Fig. \ref{fig:Bislugfig} (d) we report the steady velocity $\dot{x}_\infty$
of the bi-slug as function of $L^{-1}$ for the same
combinations of contact angles. We observe that $\dot{x}_\infty\simeq v_\infty$
(dotted lines) in the limit of long bi-slugs,
while, as the bi-slugs shorten, $\dot{x}_\infty$levels off,
implying the importance of additional channels for
energy dissipation.

To assess whether in our numerical model the additional dissipation 
originates predominantly from the viscous bending of the fluid interfaces,
we measure dynamic angles for all the fluid interfaces, fitting the fluid
interfacial profiles with circles \cite{Kusumaatmaja2016}. 
In Fig. \ref{fig:Bislugfig} (e) we report the contact
angle difference $\Delta\theta=\theta(\dot{x}_\infty)-\theta(0)$,
and observe a linear dependence with the bi-slug speed $\dot{x}_\infty$.

Motivated by this observation, we perform linear fits and introduce
the correction $\Delta\theta$ in the evaluation surface tension
unbalance $\Delta \theta$, Eqn. \eqref{equ:capillaryforce}.
The corrected model is depicted by solid lines in Fig. \ref{fig:Bislugfig} (d),
and shows excellent agreement with the measured values of $\dot{x}_\infty$.

\subsection{Contact line slip}

We now further employ the numerical experiment of self-propelled
bi-slugs to quantify the slip properties of our ternary model.
While a similar analysis could be carried out also for the capillary filling,
the bi-slug geometry has the advantage that trains of drops approach a steady 
motion with constant velocity, which can be measure  more accurately.
Furthermore, by tuning the length of the bi-slugs it is possible to vary
accurately the velocity in a wide range.

As shown by Briant \cite{Briant2004b,Briant2004a}, in multiphase Lattice Boltzmann
models, the main slip mechanism relies on evaporation-condensation of the 
liquid interface,  while in multicomponent models the contact line advances 
in virtue of the diffusion of the phase field \cite{JACQMIN2000,Kusumaatmaja2016}. 
When coupling multiphase and multicomponent models, both evaporation/condensation 
and diffusion mechanisms occur at the liquid-gas interface. 
In contrast, at the liquid-liquid interface, only the diffusion mechanism 
is important, as the density $\rho$ does not vary.

Following Cox's analysis \cite{COX1986,COX1998}, the viscous bending of a fluid
interface is described by
\begin{equation}
\label{equ:Coxapparent}
g(\theta,\lambda)-g(\theta_w,\lambda)={\rm Ca}\ln(L_c/L_s),
\end{equation}
where $\theta$ is a dynamic contact angle measured at a macroscopic distance from the surface,
and $\theta_w$ is the equilibrium contact angle at the solid boundaries. 
The Capillary number ${\rm Ca}=\mu \dot{x}_\infty /\gamma$ represents the
non-dimensional velocity of the interface, where the viscosity $\mu=\mu_{\rm adv}$ is 
referred to the invading fluid.
In our simulation we identify the macroscopic distance $L_c$ 
with the channel height $H$, and interpret the microscopic length $L_s$ 
as an estimate for the slip length. The parameter $\lambda=\mu_{\rm adv}/\mu_{\rm rec}$ 
describes the ratio between the dynamic viscosity of the invading $\mu_{\rm adv}$ 
and resisting $\mu_{\rm res}$ fluids. 

For liquid with equal density we have $\lambda=\nu_{\rm adv}/\nu_{\rm rec}$. 
Specifically, for the bi-slug simulations $\lambda=1$ 
at the liquid-liquid interface,  $\lambda \simeq 10^3$ for a liquid displacing 
the gas phase and $\lambda \simeq 10^{-3}$ for the gas displacing a liquid phase. 
The function $g(\theta,\lambda)=\int_0^\theta 1/f(\phi,\lambda)d\phi$
is a known function of $\theta$ and $\lambda$, given 
in Refs. \cite{COX1986} and \cite{COX1998}:
\begin{widetext}
\begin{equation}
\label{equ:Coxlful}
f(\theta)=\frac{2\sin\theta(\lambda^2(\theta^2-\sin^2\theta)+2\lambda(\theta(\pi-\theta)
+\sin^2\theta)+((\pi-\theta)^2-\sin^2\theta))}
{\lambda(\theta^2-\sin^2\theta)((\pi-\theta)+\sin\theta\cos\theta)+((\pi-\theta)^2
-\sin^2\theta)(\theta-\sin\theta\cos\theta)}.
\end{equation}
\end{widetext}

To systematically explore the slip properties, we perform simulations for
two sets of contact angles. In the first set
we fix $\theta_{23}=60^\circ$ and vary systematically $\theta_{21}=\theta_{31}$
in the range $[50^\circ,120^\circ]$. In the second set we fix 
$\theta_{21}=\theta_{31}=90^\circ$ and vary systematically $\theta_{23}$
in the range $[30^\circ,150^\circ]$. The first set allows us to extract 
information for the liquid-gas interfaces, while the second set
for the liquid-liquid interface.

For each combination of contact angles we simulate the motion of bi-slugs 
for a wide range of lengths and speeds. Furthermore we compute
the capillary length ${\rm Ca}$ of the advancing fluid (which can be 
either a liquid or the gas phase, depending the interface considered),
and evaluate the Cox function $g(\theta)$ in Eqn. \eqref{equ:Coxapparent} 
for the appropriate value of viscosity contrast $\lambda$. 
Due to the limited variation of the dynamic contact angles 
(in a range of a few degrees) for simplicity we  perform a linear 
regression to evaluate the slope $m=\partial g(\theta)/\partial {\rm Ca}=\ln(L_c/L_s)$. 
Finally, introducing the geometric parameter $L_c=H=39$, 
we estimate the slip length as $L_s=L_c\exp(-m)$.

\begin{figure}[tb]
\centering
\includegraphics[width=0.7\columnwidth]{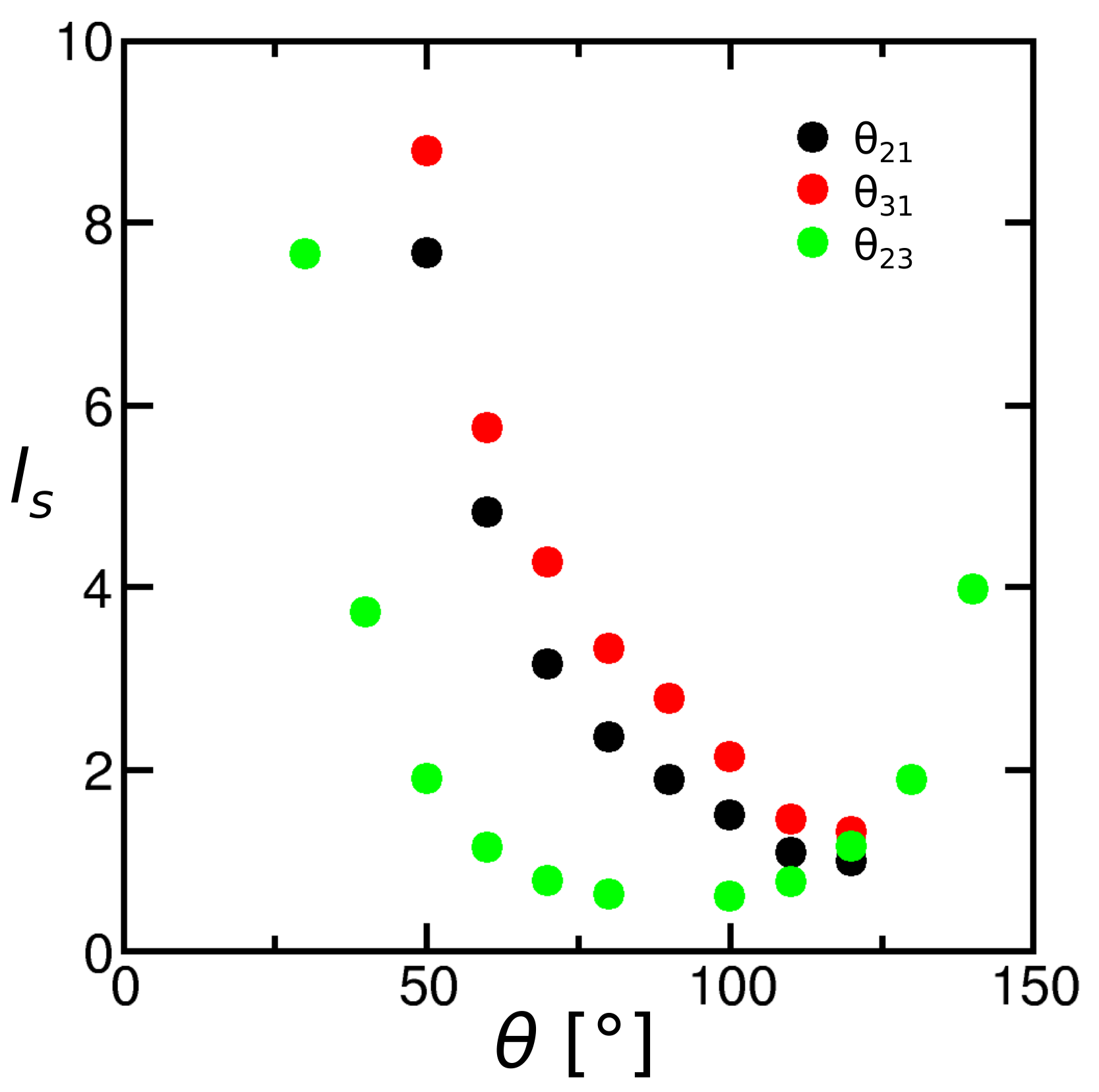}
\caption{
\label{fig:sliplenght} 
(Color online)
Estimated slip length vs. equilibrium contact angle for the liquid-liquid
and the liquid-gas interfaces. }	
\end{figure}

In Fig. \ref{fig:sliplenght}, we compare $L_s$ for the three
interfaces, as function of the equilibrium contact angle.
More specifically, for our geometry we obtain $\theta_{21}$ (receding),  
$\theta_{31}$ (advancing) and  $\theta_{23}$ (advancing for 
$\theta_{23}<90^\circ$ and receding for $\theta_{23}>90^\circ$).
The slip length for the liquid-liquid interface shows a minimum for
$\theta_{23}=90^\circ$ (the data point not present, because for this 
combination of angles we have no net driving force $\Delta \gamma=0$) 
and increases symmetrically for larger and smaller angles. 
In contrast the slip length for the liquid-gas
interfaces shows a monotonic decrease as the equilibrium contact angle
increases. For the last available data point, at $\theta=120^\circ$,
$L_s$ is similar for all three interfaces, while for smaller angles
$L_s$ is significantly larger for the liquid-gas interfaces.

These results show that the slip properties in the system strongly depend on 
the nature of the fluid-fluid interface. In our tests the liquid-gas 
interfaces present a larger slip length (up to a factor 4) compared to
the liquid-liquid interface, likely due to the combined effect of 
two distinct mechanisms operating on the density $\rho$ and the field $\phi$.


\section{Discussion and conclusions}
\label{conclusions}
	
In this work we have thoroughly benchmarked our ternary high density ratio
free energy model, and provided guidelines to select the free energy 
parameters for obtaining a wide range of surface tension combinations. 
We have quantified the deviations of the interface profile by measuring the 
\enquote{Deformation coefficient}, and systematically investigated 8 subspaces, covering 
relevant combinations of parameters. All data are reported in the supplementary 
information, including fitting functions to estimate the surface tensions. 

We have compared three methods for wetting of solid boundaries, namely
\emph{force}, \emph{geometric extrapolation} and \emph{geometric interpolation}.
Of the two geometric methods, \emph{geometric interpolation} is significantly 
more accurate. The \emph{force} method provides an useful alternative to 
geometric methods, as it does not require us to detect the fluid interface 
a priori, and automatically satisfies the Girifalco-Good relation, 
Eqn. \eqref{equ:GirifalcoGood}. 

The benchmark on the dynamics of capillary filling shows that 
the force method is slightly less accurate than the geometric methods.
The deviations in the measured pre-factor of the Washburn law increase
as the material contact angles depart from neutral wetting. 
Because the additional force terms also increase, we expect
that the deviations are related to additional spurious velocities 
generated by the forcing terms. At the same time no spurious
velocities are observed in the geometric methods.

Furthermore we have have performed a ternary specific benchmark, and
studied the motion of self-propelled bi-slugs forming three finite
and unbalanced contact angles. The analytic model
for the bi-slug motion, derived by assuming equilibrium values for the contact
angles, accurately captures the linear dependence of steady state velocities
from the inverse of the bi-slug length, for long trains of drops
and small velocities. The level-off of the velocity experimentally observed for shorter bi-slugs is captured in our simulations by accounting for the dynamic angle correction in the driving force. The agreement shows that the viscous bending of the liquid interfaces represents the main correction in the model.  

Finally we have shown that coupling multiphase and multicomponent
models leads to significant differences in the
slip properties of liquid-liquid and liquid-gas interfaces.
While for liquid-liquid interfaces the only slip mechanism is related
to the diffusion of the field $\phi$, for liquid-gas interfaces
the slip mechanism combines the diffusion of the field $\phi$ and
the evaporation/condensation of the density $\rho$.  

In our tests, at parity of interfacial properties, the slip length 
varies with the material contact angle and is generally larger for 
the liquid-gas interfaces than for the liquid-liquid interfaces. 
A more detailed analysis of the slip properties of the system 
will be the subject of a future investigation.


\section{Acknowledgement}
\label{Acknowledgement}

NB gratefully acknowledges financial support from University of Northumbria 
at Newcastle via a Postgraduate Research Studentship. 
IK acknowledges support by European Research Council (ERC) Advanced Grant 834763-PonD 
and by Swiss National Science Foundation (SNF) grant 200021\_172640.
HK thanks Procter \& Gamble (P\&G) and EPSRC for funding (EP/P007139/1).
CS acknowledges support from Northumbria University through the Vice-Chancellor's 
Fellowship Programme.

\bibliographystyle{apsrev4-1}
\bibliography{references}
	
\end{document}